\documentclass[fleqn,usenatbib]{mnras}

\usepackage{newtxtext,newtxmath}
\usepackage[T1]{fontenc}

\DeclareRobustCommand{\VAN}[3]{#2}
\let\VANthebibliography\thebibliography
\def\thebibliography{\DeclareRobustCommand{\VAN}[3]{##3}\VANthebibliography}

\usepackage{amsmath}
\usepackage{graphicx}

\usepackage{soul}
\usepackage{siunitx}
\usepackage{tikz}
\usetikzlibrary{shapes.geometric,arrows}
\usetikzlibrary{graphs}

\tikzstyle{app} = [rectangle,rounded corners,minimum width=3cm,%
    minimum height = 1cm, text centered, draw =black,fill=red!30]
\tikzstyle{class} = [rectangle,rounded corners,minimum width=2cm,%
    minimum height = .8cm, text centered, draw =black,fill=white]
\tikzstyle{function} = [minimum width=2cm,%
    minimum height = .8cm, text centered, draw =none]
\tikzstyle{arrow}=[thick,->,>=stealth]
\tikzstyle{arrowv}=[thick,dashed,->,>=stealth]
\newcommand{\order}{\mathcal{O}}

\DeclareMathOperator{\erfc}{erfc}

\newcommand{\daa}{\mathcal{H}}
\newcommand{\pdaa}{\daa}

\newcommand{\GR}{GR}
\newcommand{\PM}{PM}

\newcommand{\FLRW}{FLRW}

\newcommand{\gadget}{\textsc{\hbox{Gadget-4}}}
\newcommand{\grgadget}{\textsc{GrGadget}}
\newcommand{\libgevolution}{\textsc{Libgevolution}}
\newcommand{\latfield}{\textsc{LATfield}}
\newcommand{\gramses}{\textsc{Gramses}}

\newcommand{\gevolution}{\textsc{Gevolution}}
\newcommand{\mpi}{\textsc{MPI}}
\newcommand{\pmesh}{\textsc{PM}}
\newcommand{\treepm}{Tree\textsc{PM}}

\newcommand{\testsmall}{\texttt{N64}}
\newcommand{\testmed}{\texttt{N256}}
\newcommand{\testbig}{\texttt{high\_res}}

\title[{\grgadget}]{{\grgadget}: an N-body TreePM relativistic code for cosmological simulations}

\author[E. Quintana-Miranda et al.]{
Eduardo Quintana-Miranda,$^{1,2,3}$\thanks{E-mail:
eduardo.quintanamiranda@phd.units.it}
Pierluigi Monaco$^{1,2,3,4}$
and Luca Tornatore$^{1,2,3}$
\\
$^1$ Dipartimento di Fisica, Sezione di Astronomia, via G.B. Tiepolo 11, I-34143 Trieste, Italy\\
$^2$ INAF -- Istituto Nazionale di Astrofisica, Osservatorio Astronomico di Trieste, via G.B. Tiepolo 11, I-34143 Trieste, Italy\\
$^3$ IFPU -- Institute for the Fundamental Physics of the Universe, via Beirut 2, I-34100 Trieste, Italy\\
$^4$ INFN -- Istituto Nazionale di Fisica Nucleare, Via Valerio 2, I-34127 Trieste, Italy}

\date{Accepted XXX. Received YYY; in original form ZZZ}

\pubyear{2022}

\begin{document}
\label{firstpage}
\pagerange{\pageref{firstpage}--\pageref{lastpage}}
\maketitle

\begin{abstract}
We present the merging of the Particle-Mesh (PM) relativistic {\gevolution} code with the TreePM {\gadget} code, 
%that offers a TreePM relativistic code suited to 
with the aim of studying general relativity effects in cosmology. Our code, called {\grgadget}, is able to track the evolution of metric
perturbations in the weak field limit by using {\gevolution}'s implementation of a relativistic PM in the Poisson gauge.
To achieve this, starting from {\gevolution} we have written a C++ library called {\libgevolution},
that allows a code to %programmatically 
access and use the same abstractions and resources that {\gevolution} uses for its PM-only N-body simulations.
The code works under the assumption that 
%in a matter-dominated universe 
particle interactions at short distances can be approximated as Newtonian, so that we can combine the forces computed with a Newtonian Tree with those computed with a relativistic PM.
The result is a TreePM simulation code that represents 
%gravity as the effect of 
metric perturbations at the scales where they are relevant,
%the Newtonian limit is not accurate, 
while resolving non-linear structures. 
%The code can be used to obtain on-the-fly the information of the metric
%perturbations that can be used to simulate realistic GR effects produced mainly by
%perturbations of light's geodesics.
We validate our code by closely matching {\gadget} forces, computed with the Tree switched off, with those computed with {\libgevolution } in the Newtonian limit. 
With {\grgadget} we obtain a matter power spectrum that is compatible with Newtonian {\gadget} at small scales and contains GR features at large scales that are consistent with results
%the simulations 
obtained with {\gevolution}.
We demonstrate that, due to the better resolution of the highly non-linear regime,  the representation of the relativistic fields sampled on the mesh improves with respect to the PM-only simulations.
%Due to a much better resolution of high non-linearities, we notice, with respect
%to PM-only simulations, an improvement in the representation of the relativistic
%fields on scales near the Nyquist frequency limit imposed by the PM.

% This is a simple template for authors to write new MNRAS papers.
% The abstract should briefly describe the aims, methods, and main results of the paper.
% It should be a single paragraph not more than 250 words (200 words for Letters).
% No references should appear in the abstract.
\end{abstract}

\begin{keywords}
cosmology: theory -- large-scale structure of the Universe
\end{keywords}

%%%%%%%%%%%%%%%%%%%%%%%%%%%%%%%%%%%%%%%%%%%%%%%%%%

%%%%%%%%%%%%%%%%% BODY OF PAPER %%%%%%%%%%%%%%%%%%

% All papers should start with an Introduction section, which sets the work
% in context, cites relevant earlier studies in the field by \citet{Fournier1901},
% and describes the problem the authors aim to solve \citep[e.g.][]{vanDijk1902}.
% Multiple citations can be joined in a simple way like \citet{deLaguarde1903, delaGuarde1904}.
\section{Introduction}
% \eduardo{The context: DM, DE, Mod. gravity and testing GR.}
\noindent 

The state of the art of precision cosmology provides a standard cosmological model, $\Lambda$CDM, that is consistent with most observational evidence on large scales, but relies on the existence of a dark sector populated by Dark Matter (DM) and Dark Energy (DE).
The first is responsible for the formation of cosmological structures such as galaxies and their large-scale density field, while the second causes the observed accelerated expansion of the universe in the present epoch. 
Their physical nature is an open problem, since the only evidence of their existence comes from their gravitational interaction with visible matter.
A possible explanation is that the dark sector is due to a misrepresentation of gravity, that on large scales does not follow Einstein's General Relativity (\GR), at the basis of the $\Lambda$CDM model.

This fact has triggered a wave of interest in modifications of GR, that can lead to extra terms that explain dark energy or dark matter \citep[see, e.g.,][and references therein]{Silvestri2009,Capozziello2012}.
Such modifications must be significant only on large scales or low density, because {\GR} is very accurate in predicting planetary orbits, light deflection and Doppler effects in solar
system tests and has more recently been successfully tested with the detection of
gravitational waves \citep{LIGO2016} and the direct imaging of black hole event horizons \citep{EHT2019}.

In order to characterize dark energy in the age of its dominance, many projects
have been planned to survey large parts of the sky and probe the large-scale
distribution of matter using galaxy clustering and galaxy lensing, both from the
ground (DES\footnote{\url{www.darkenergysurvey.org}}, \citealt{des2017};
DESI\footnote{\url{www.desi.lbl.gov}}, \citealt{desi};
Rubin's LSST\footnote{\url{www.lsst.org}}, \citealt{lsst2019};
SKAO\footnote{\url{www.skao.int}} surveys) %, \citealt{})
and from space
(Euclid\footnote{\url{sci.esa.int/web/euclid}}, \citealt{euclid_redbook};
Roman\footnote{\url{roman.gsfc.nasa.gov}}, \citealt{roman2015};
SphereX\footnote{\url{spherex.caltech.edu}}, \citealt{spherex2014}).
Some of these surveys have already started to produce a flood of data that will soon lead to a
precise characterization of the galaxy and matter density fields. A comparison
of these observations to model predictions, either using summary statistics or
field-level inference, will lead to unprecedented tests not only of the
cosmological model but also of the gravity theory behind it.
%\eduardo{The observational landscape with higher precision and the role of simulations. I could citet LSST 1201.2434, Euclid 1110.3193, DESI 1611.00036.}
With precision being guaranteed by the amount of available high-quality data, accuracy will be achieved only by rigorous control of systematics, both in the data and in theory predictions.

%\eduardo{The need for GR codes over Newtonian. Alam 2020, Gevolution Paper and
%GRAMSES paper.}

The highly non-linear nature of the observed density field and the non-locality of gravity make cosmological simulations necessary to 
compare the predictions of current theories with the observations at an increasing level of accuracy.
Yet, most 
of the widely adopted
simulation codes, like e.g. \gadget\ \citep{Springel2021}, use Newtonian dynamics for the evolution of matter perturbations. This is not the ideal configuration to pass from the unobservable distribution of matter in a periodic comoving box to the observable distribution of light in the past light cone. 
Relativistic corrections can be added \emph{a posteriori} by post-processing Newtonian simulation outputs; one specific example of this approach is the modeling of lensing due to the distortion of null geodesics \citep{BartelmannSchneider2001}, while a more comprehensive approach to adding relativistic effects is presented by \citet{Bertacca2017}.
However, even though the biases introduced by this approach are expected to be small, a fully self-consistent approach is necessary to convincingly demonstrate our ability of controlling theory systematics. For instance, galaxy clustering is affected by magnification bias due to lensing, and neglecting this effect induces a non-negligible bias in parameter estimation \citep{Lepori2020,Alam2021}.
This is even more true when modified gravity theories are used: extensions of gravity are typically derived in a full relativistic context, and while they influence the Newtonian limit of gravity, the small but measurable relativistic effects may provide smoking-gun signals of a specific class of gravity theories. In this sense, restricting to the treatment of the Newtonian limit of modified gravity theories \citep[as, e.g., in][]{Puchwein2013} may leave out crucial observable signatures.

% \eduardo{The landscape of current GR codes and their limitations.}
Two examples of fully relativistic N-body codes for the evolution of cosmic perturbations, that integrate Einstein's equations to follow the motion of massive particles along their geodesics, are the Adaptive Mesh Refinement (AMR) code \gramses\ \citep{Hinojosa2020} and the Particle-Mesh (PM) code \gevolution\ \citep{Adamek2016}.
These have proven to be precious tools to produce accurate cosmological predictions, like a self-consistent treatment of massive neutrinos \citep{Adamek2022}, and to explore phenomena that were previously overlooked, like the strength of the frame dragging field
acting on dark matter haloes \citep{barrerahinojosa2020vector}.
These codes sample the fields in a mesh that fills the simulated volume, but while {\gramses} uses an AMR scheme to increase resolution only where it is needed, PM schemes working on a single non-adaptive mesh are well known to be limited by memory, so they are unable to achieve the large dynamic range required, e.g., to resolve DM halos in large cosmological volumes.
The integration of Newtonian particle trajectories has historically been addressed
with the introduction of an oct-tree data structure \citep{BarnesHut1986}, that provides a $N\log N$ scaling for the computation of gravity without compromising its accuracy. Because the integration of large-scale perturbations is very slow in this scheme, such an oct-tree is used to compute short-range forces, and is complemented by a Particle-Mesh (\PM) code on large scales. The resulting algorithm is commonly called TreePM, and it is the standard gravity solver for {\gadget}.

As we will show in next Section, deviations from a pure Newtonian approach become significant on scales that are comparable with the Hubble horizon, so a Newtonian treatment of small-scale clustering, performed by the Tree algorithm, would introduce a negligible error if large scales are treated by a fully relativistic gravity solver. This can be achieved, in a TreePM scheme, by using a relativistic PM code for large-scale gravity, where relativistic potentials are sampled on a small enough mesh so as to be effectively Newtonian on the scales where the Tree code gets in.

In this paper we present an implementation of {\gadget} that uses a {\pmesh} library, based on \gevolution\ relativistic code, as the PM part of the TreePM solver. This is a step toward the construction of an ecosystem of codes and post-processing tools to perform end-to-end simulations of future surveys, with the aim of achieving optimal control of all systematics, including theoretical ones. The paper is organized as follows: Section 2 gives an overview of the theory of relativistic perturbations, with a focus on the approach used in {\gevolution}. Section 3 gives a description of the {\gadget} and {\gevolution} codes, and describes the implementation of {\libgevolution} and {\grgadget}.
Section 4 presents the tests performed to validate {\grgadget}, while Section 5 gives our conclusions.

\section{Theory of relativistic perturbations}\label{sec:theory}
% \eduardo{Theory: the metric and gauge.}
%    \eduardo{Cite 0504097, 1405.1418, 1407.7040 and Millenium.}

The success of Newtonian simulations in describing the large-scale structure of
the universe follows from the fact that, for an observer at rest with respect to
the CMB, the metric of spacetime is very close to
Friedmann-Lemaitre-Robertson-Walker's (FLRW). Deviations from the Newtonian approach are expected to be
significant, albeit small, on scales near the Hubble horizon, or when the
energy-momentum tensor has relativistic components like radiation or fast
massive neutrinos. Deviations from FLRW metric are expected to be strong in the
proximity of compact objects, but this happens on scales that are far smaller
than the resolution that can be afforded in simulations of large comoving
volumes. It is thus fair to assume that the perturbations to the metric are
small and can be described in a weak-field regime. This does not imply that
deviations of the components of the energy-momentum tensor from homogeneity are
assumed to be small, density perturbations can be highly non-linear: what we
require is that the size of self-gravitating objects is much larger than their
gravitational radius.

The \gevolution\ code \citep[see][]{Adamek2016} models the spacetime metric with
a perturbed FLRW metric in the weak field regime.
%assuming those perturbations are small so that only linear and some second
%order terms are preserved in the equations of motion.
In the \emph{Poisson gauge} the metric can be written as:
\begin{equation}
\begin{split}
    ds^2 = & a^2 \Big( -c^2\, d\tau^2 (1 + 2\Psi) - 2 c\, d\tau dx^i B_i
    + \\ & + dx^i dx^j \big(\gamma_{ij}(1-2\Phi)  
    + h_{ij}
    \big)\Big),
\end{split}
    \label{eq:metric_poisson_gauge}
\end{equation}
where $a(\tau)$ is the scale factor of the FLRW background, $\tau$ is the
conformal time and $x^i$ are the space coordinates. It is possible to exploit the residual degrees of freedom of the metric to impose 
%to the fields $\Psi$, $\Phi$, $B_i$ and $h_{ij}$ 
%gauge fixing 
the conditions $B_{i|}{}^{i}=0$,
$h_{i}{}^i=0$ and $h_{ij|}{}^j=0$.
In our notation, repeated latin indexes denote Einstein's summation over the
spatial coordinates $1,2,3$ and the vertical bar subscript, e.g. $B_{i|j}$, denotes a
covariant derivative with respect to the affine connection that emerges from the
background spatial metric $\gamma_{ij}$. 

The choice of the Poisson gauge is convenient because the two potentials $\Psi$ and $\Phi$
are the gauge-invariant Bardeen potentials, and in the Newtonian limit the
the field $\Psi$ can be interpreted as the gravitational potential.
In other words, this is the gauge in which the standard N-body solver is
integrating the right equations of motion in the Newtonian limit
\citep{Chisari_2011}.

\subsection{Field equations}
The background, characterized by $a(\tau)$, is by construction a solution of the
Einstein's equations in the presence of a homogeneous and isotropic
energy-momentum tensor $\bar T^{\mu}{}_{\nu}$:
\begin{equation}
    \bar G^{\mu}{}_{\nu} = - \frac{8\pi G}{c^4} \bar T^{\mu}{}_{\nu}\, ,
    \label{eq:einstein_background}
\end{equation}
where $\bar G^{\mu}{}_{\nu}$ is Einstein's tensor constructed from the metric
(\ref{eq:metric_poisson_gauge}) with the perturbations $\Psi,\Phi,B_i,h_{ij}$ set
to zero. Applying equation (\ref{eq:einstein_background}) to the FLRW metric one obtains
Friedmann's equations.

%The full Einstein's equations instead provide a means for computing the
To solve for the perturbations of the metric, the usual procedure 
%to separate background constraints from the evolution of the perturbations, 
consist in subtracting
(\ref{eq:einstein_background}) from the full Einstein's equations:
\begin{equation}
    G^{\mu}{}_{\nu} - \bar G^{\mu}{}_{\nu} 
    = - \frac{8\pi G}{c^4}(T^{\mu}{}_{\nu} -\bar T^{\mu}{}_{\nu})\, .
    \label{eq:einstein_perturbation}
\end{equation}
The right hand side now contains the perturbation of the energy-momentum
tensor due to inhomogeneities in mass and energy distributions, while the left
hand side is a very complicated non-linear expression containing the potentials
$\Psi,\Phi,B_i,h_{ij}$ and their space-time derivatives up to second order.

To reach a tractable set of equations that we can interpret and solve
numerically, we apply the weak field assumption. The perturbations
$\Psi,\Phi,B_i,h_{ij}$ are assumed to be of order $\epsilon \ll 1$.
Spatial derivatives are known to increase their amplitude by a factor
of $\epsilon^{-1/2}$, accounting for the presence of shortwave fluctuations induced
by the non-linear structure in the energy-momentum tensor, while time derivatives are assumed to preserve the perturbation order.
Then one can expand $G^{\mu}{}_{\nu}-\bar G^{\mu}{}_{\nu}$ in terms of the
metric perturbations, neglecting contributions with order higher than
$\epsilon$. For example: $\Phi$ is a term of order $\order(\epsilon)$,
$\Phi_{,i}$ has order $\order(\epsilon^{1/2})$, %(the derivative increases the amplitude),
$\Phi_{|n}{}^{n}$ is a leading term (order 1, because of the second derivative),
quadratic terms like $\Phi_{,n} \Phi{,}{}^{n}$ are $\order(\epsilon)$,
and a term like $\Phi_{,00}$ is considered as $\order(\epsilon)$.
%because time derivative are not considered to increase the overall amplitude of the fields. 
This type of expansion is known as the \emph{shortwave correction} \citep{adamek2014}.

Furthermore, experience has shown that the scalar perturbations $\Phi$ and
$\Psi$ are generally larger than the vector and tensor perturbations $B_i$ and
$h_{ij}$. Indeed, the scalar potentials, that are sourced by the density perturbation $\Delta T^{00}$,  become the Newtonian potential in the Newtonian limit, while the vector perturbation $B_i$ is sourced by $\Delta T^{0i}$, that is small by a factor of $v/c$ for non-relativistic matter perturbations, and $h_{ij}$ by $\Delta T^{ij}$, that is suppressed by a $(v/c)^2$ factor. 
Hence, it is fair to drop quadratic terms of $B_i$ and $h_{ij}$ in this weak field limit approximation.

In this approximation, from Eq. (\ref{eq:einstein_perturbation}) it descends
that its time-time component yields a Poisson-like equation for the scalar $\Phi$:
\begin{equation}
\begin{split}
    \Phi_{|n}{}^n (1 + 4\Phi)
    & - 3 \frac{\daa}{c^2}\Phi_{,0}
    + 3 \frac{\pdaa^2}{c^2}(\chi - \Phi)
    + \frac{3}{2}\Phi_{|n}\Phi_{|}{}^n
    % + \frac{\Phi}{2} \bRicci^{(3)}
    % - \frac{h^{nl}}{2}(\Phi_{|nl}-\Psi_{|nl})
    % - \frac{1}{4} h^l{}_n \bRicci^{(3)}{}^n{}_l
    \\& =  \frac{4\pi G a^2}{c^4} \Delta T^{0}{}_{0} \, ,
\end{split}
    \label{eq:einstein_perturbations_tt_gev}
\end{equation}
where $\daa = a^{-1}\frac{da}{d\tau}$ and $\chi = \Phi - \Psi$.
From the time-space section of eq. (\ref{eq:einstein_perturbation}) we
obtain:
\begin{equation}
     - \frac{ B_{i|n}{}^n }{4c}
     - \frac{\Phi_{,i0}}{c^2}
     - \frac{\daa}{c^2} (\Phi_{,i} - \chi_{,i})
     %-\frac{B_n}{2c} \bRicci^{(3)n}{}_i
     %- \frac{B_i}{2c} \Phi_{|n}{}^n
     %- \frac{B^n}{2c} \Phi_{|ni}
     =  -\frac{4\pi G a^2}{c^4} \Delta T^{0}{}_{i} \, ,
    % \label{eq:einstein_perturbations_ts_gev_full}
\end{equation}
that, taking advantage of the %gauge 
condition $B_{n|}{}^n = 0$, can be reduced to:
\begin{equation}
     - \frac{ B_{i|n}{}^n }{4c}
     =  -\frac{4\pi G a^2}{c^4} P_{\perp}\Delta T^{0}{}_{i} \, ,
    \label{eq:einstein_perturbations_ts_gev}
\end{equation}
where $P_{\perp}$ is a linear operator that selects from a vector field its
divergenceless component.

The traceless part of the spatial section of eq. \ref{eq:einstein_perturbation} 
leads to:
\begin{equation}
\begin{split}
     & 
     \left(
        \delta^j{}_b \delta^a{}_i-\frac{1}{3} \delta^a{}_b \delta^j{}_i \right) 
     \Bigg[
        \chi_{|j}{}^i
        -2 \Phi_{|j}{}^i \chi
        + 4\Phi\Phi_{|j}{}^i
        + 2\Phi_{|j}\Phi_{|}{}^i
        \\&\qquad
        + \frac{1}{2c^2}h^i{}_{j,00}
        + \frac{\daa}{c^2} h^i{}_{j,0}
        - \frac{1}{2} h^i{}_{j|n}{}^n
        \\&\qquad
        + \frac{1}{2c}\left(\frac{\partial}{\partial \tau} + 2\daa\right)\left(B^i{}_{|j}+B_{j|}{}^i\right)
     \Bigg]
     \\&=
     \left( \delta^j{}_b \delta^a{}_i-\frac{1}{3} \delta^a{}_b \delta^j{}_i \right) 
     \left( - \frac{8\pi G}{c^4} \Delta T^{i}{}_{j} \right)\, ,
\end{split}
    \label{eq:einstein_perturbations_ss_traceless}
\end{equation}
from which we can determine the rest of the metric degrees of freedom 
$\chi$ and $h_{ij}$. Since the source of $\chi$ and $h_{ij}$ are the perturbation of the
of the energy-momentum tensor $\Delta T^{i}{}_{j}$, their amplitude in a
matter dominated universe is suppressed by a factor $(v/c)^2$. That is equivalent
to say: since dark matter is non-relativistic, $\chi$ and $h_{ij}$ must be very small with
respect to $\Phi$ or even $B_i$.
%For this reason quadratic terms involving $\chi$ are neglected.

%\begin{equation}
%    \begin{split}
%    -\chi_{|n}{}^n
%    &+9\frac{\daa}{c^2}\Phi_{,0}
%    +3\frac{1}{c^2}\Phi_{,00}
%    +6\frac{\daa_{,0}}{c^2}\Phi
%    +3\frac{\daa^2}{c^2}\Phi
%    \\&= 
%    -\frac{4\pi G a^2}{c^4} \Delta T^{n}{}_{n} \, .
%    \end{split}
%    \label{eq:einstein_perturbations_ss_gev_trace}
%\end{equation}

%On the other hand, 
%Finally the traceless components of eq. (\ref{eq:einstein_perturbation}) yield a set of equations for $h_{ij}$:
%\begin{equation}
%    \begin{split}
%        \frac{h^{i}{}_{j,00}}{2c^2} - \frac{h^{i}{}_{j|n}{}^n}{2} 
%        + \frac{\daa}{c^2} h^{i}{}_{j,0}
%    = 
%    -\frac{8\pi G a^2}{c^4} P_{\perp}\left( \Delta T^{i}{}_{j} - \frac{\delta^i{}_{j}}{3}
%    \Delta T^n{}_{n} \right) \, .
%    \end{split}
%    \label{eq:einstein_perturbations_ss_gev_traceless}
%\end{equation}
%However, the effect of tensor modes is very small, we will not consider them in this paper.

As a matter of fact, V1.2 of {\gevolution} implements an improved expansion of the metric perturbations, that has been presented in \cite{Adamek2017}. For our tests we used the implementation of the original expansion, the one presented above. However, the improved expansion has been ported to {\libgevolution} and will be used when analysing result on the past light cone. We do not expect the results presented in this paper to depend on the specific expansion used.

%\eduardo{Geodesics}
\subsection{Motion of particles along geodesics}
    \newcommand{\qq}{q^2+m^2 a^2}
Massive particles move along geodesics, whose equation can be expressed as:
{\newcommand{\sroot}{\sqrt{(mca)^2 + p^2}}
\begin{equation}
    \begin{split}
    \frac{dx^i}{d\tau}
    = 
    & \frac{c p^i}{\sroot}
    + c B^i
    %- \frac{c h^{ni} p_n}{\sroot}
    %+\\&\qquad
    \\& + \frac{c p^i}{\sroot}
    \left(
        \Psi
        + \Phi \frac{2(mac)^2+p^2}{(mac)^2+p^2}
        %+ \frac{h_{nm} p^n p^m}{2((mac)^2+p^2)}
    \right)\, ,
    \end{split}
    \label{eq:hamilton_eq1}
\end{equation}
\begin{equation}
    \begin{split}
    \frac{dp_i}{d\tau}=
    &
    -c\Big(
        p^n B_{n|i}
        + \Psi_{,i}\sroot
     + \frac{p^2 \Phi_{,i}}{\sroot}
        %- \frac{p^np^m h_{nm,i}}{2\sroot}
    \Big)\, ,
    \end{split}
    \label{eq:hamilton_eq2}
\end{equation}
}

\noindent
where $p^i$ is the space part of the particle momentum and $p$ its norm.
The right hand side in the last equation is the generalized force acting on the particles.
The term proportional to $\Psi_{,i}$ becomes the Newtonian force in the limit of small velocities,
while $p^n B_{n|i}$ represent the corrections due to \emph{frame dragging} and the third term in parenthesis is a further relativistic correction.
     
%\eduardo{Energy-momentum tensor.}
The energy-momentum tensor is constructed from the knowledge of particle
positions and momenta, but its computation depends on the perturbed metric.
This means that, in Eqs. (\ref{eq:einstein_perturbations_tt_gev}), 
(\ref{eq:einstein_perturbations_ts_gev}),
and (\ref{eq:einstein_perturbations_ss_traceless}), 
the source terms on the right hand sides depend on the potentials themselves. These implicit equations may be solved starting from the potentials at the previous time step and solving the equations iteratively until convergence. 
The integration scheme that {\gevolution} implements is simpler: at each time step the energy momentum tensor is computed using the potentials from the previous step, then the Poisson equations
are solved to find the updated potentials, that will be used in the next time step to compute the energy-momentum tensor.

%\eduardo{Newtonian Limit.}
The Newtonian limit is recovered when we consider Fourier modes
larger than $\daa/c$ and we further neglect $B_i$ and consider $\Phi\ll 1$;
then equation (\ref{eq:einstein_perturbations_tt_gev}) becomes:
\begin{equation}
\begin{split}
    \Phi_{|n}{}^n 
     =  \frac{4\pi G a^2}{c^4} \Delta T^{0}{}_{0}
\end{split}
    \label{eq:einstein_perturbations_tt_gev_newtonian}
\end{equation}
while (\ref{eq:hamilton_eq1}) and (\ref{eq:hamilton_eq2}) become:
\begin{equation}
    \begin{split}
    \frac{dx^i}{d\tau}
    = 
    & \frac{p^i}{ma}\, ,
    \end{split}
    \label{eq:hamilton_eq5}
\end{equation}
\begin{equation}
    \begin{split}
    \frac{dp_i}{d\tau}=
       - \Phi_{,i} m c^2 a\, .
    \end{split}
    \label{eq:hamilton_eq6}
\end{equation}

\section{Algorithms and code infrastructure}

\subsection{\gevolution}
%\eduardo{What is gevolution?}

{\gevolution}\footnote{\url{https://github.com/gevolution-code}}
\citep{Adamek2016} is an N-body relativistic cosmological code, written in C++
and parallelized with the {\mpi} paradigm.
The physical theory behind this code has been described at length in
Section~\ref{sec:theory}.
%\eduardo{Gevolution is relativistic. Gravity is solved with a PM.}
Numerically, this code implements a {\pmesh} scheme to follow the evolution of
energy-momentum tensor perturbations. As in PM codes, the advantage of working
with a single grid and using Fast Fourier Transforms (FFTs) to solve the Poisson-like
equations for the fields is paid with a high cost in memory, of $\order(N^3)$
where $N$ is the number of grid points per dimension.

\gevolution, can run in either \emph{Newton} or \emph{General Relativity} modes.
The Newtonian gravity solver inverts the Laplace operator in the 
Poisson equation for the Newtonian potential,
Eq.~\ref{eq:einstein_perturbations_tt_gev_newtonian}.
%\[
%    \nabla^2 \Phi_N = 4\pi G a^3 \rho,
%\]
%where $\rho$ is the mass per comoving volume. 
When running the General Relativity mode, the code solves
Eqs.~\ref{eq:einstein_perturbations_tt_gev},
\ref{eq:einstein_perturbations_ts_gev} and
\ref{eq:einstein_perturbations_ss_traceless},
%and \ref{eq:einstein_perturbations_ss_gev_traceless}, 
that require the computation of the
perturbed energy-momentum tensor. This is performed using a Cloud-In-Cell (CIC)
scheme both for the density and for particle velocities; details are given in
the presentation paper.
Then the Hamiltonian forces to which particles are subjected are computed from
Eqs.~\ref{eq:hamilton_eq1} and \ref{eq:hamilton_eq2}.

\gevolution\ solves the field equations in Fourier space, using a C++ library called \latfield2 to operate FFTs on classical fields in massively parallel applications with distributed memory. \latfield2 provides a programming interface
to perform operations on the fields, either in their real or Fourier space
representations.
This library implements FFTs of 3-dimensional fields whose memory is distributed among parallel processes following a 2-dimensional
uniform decomposition of space, in which each process owns in memory a portion
of the grid with a \emph{rod} shape \citep{daverio2015latfield2}.
In this way \latfield2 overcomes the scaling limitations of a simpler
1-dimensional domain (\emph{slab}) decomposition provided by the mainstream
FFTW3 library\footnote{\url{http://fftw.org/}}.
FFTW3 is used, however, to compute 1D FFTs.

\subsection[Gadget-4]{\gadget}
%\eduardo{What is Gadget4?}
{\gadget}\footnote{\url{https://wwwmpa.mpa-garching.mpg.de/gadget4}} is a state-of-the-art TreePM N-body hydrodynamical cosmological
code written in C++ \citep[see][]{Springel2021}; it is massively parallelized in a distributed-memory paradigm using \mpi .

%\eduardo{Gadget is Newtonian. Gravity is solved with a TreePM method.}
As in most N-body codes, gravity in {\gadget} is represented in the Newtonian limit, but the equations of motion are modified to take into account the Universe expansion, obtained by integrating the Friedmann equations separately. As mentioned above, this approach is consistent with General Relativity in the Poisson gauge, and gives the 
leading-order term of weak field expansion. This amounts to  neglecting the metric degrees of freedom $B_i$, $\chi$ and $h_{ij}$, and is valid on scales much smaller than the Hubble horizon.
In a typical configuration that is convenient for large cosmological volumes, the code solves for the forces acting on each particle, representing them as the sum of two contributions, one due to the interactions with nearby particles, computed with a Tree algorithm, and one due to long-range interactions, computed with a PM algorithm.\footnote{The code can work in other configurations (a non-cosmological volume, switching off the PM, enhancing the Tree part using multipole expansion) that are however not relevant for this paper.}

%\eduardo{The Tree.}
The Tree algorithm works by partitioning the space into cubic cells, called nodes; in turn, each node is recursively partitioned into 8 children nodes %if it contains many particles, 
down to a pre-determined maximum refinement level. A tree structure tracks the list of particles that are located within each node. This structure is used to speed up the computation of gravitational force on a particle: in a particle-particle integration scheme, this force is computed by adding up a series of $\vec r\, m/r^3$ terms, one for each particle pair, but we know that the accuracy of force evaluation does not depend strongly on the small-scale distribution of distant particles, so in the Tree scheme the evaluation of gravity is performed by grouping particles that belong to the same node, under the condition that the node subtends a given aperture angle $\theta$. Particle-particle computation is then used only for the nearest neighbours.
This is equivalent to considering the leading order in a multipole expansion of the gravity force from particles belonging to a distant cell.
While the construction of the Tree is expensive in terms of computing time, it allows to achieve  $\mathcal{O}(N_p \log N_p)$ scaling for the force computation,  where $N_p$ is the total number of particles in the simulation.
Thus the Tree is able to compute with high accuracy the short wavelength modes of the gravitational interaction, while keeping the computational time low for large simulations. However, the Tree code is slow in integrating particle motions near the initial conditions, when the departures from homogeneity are small. This is why it is often coupled with a {\pmesh} code to speed up the first time steps of a cosmological box.

%\eduardo{The PM.}
The {\pmesh} algorithm represents gravity through the gravitational potential field $\Phi$, evaluated on a Cartesian cubic mesh of fixed size. The potential is found from the density field by solving the Poisson equation in Fourier space, while the force is computed from the gradient of the potential, obtained with a finite differences scheme. 
According to the Nyquist-Shannon theory, this implies that the information handled by the {\pmesh} is limited to the long modes, up to the Nyquist frequency.

\newcommand{\Phishort}{\Phi^{(S)}}
\newcommand{\Philong}{\Phi^{(L)}}
\newcommand{\tildePhishort}{\tilde\Phi^{(S)}}
\newcommand{\tildePhilong}{\tilde\Phi^{(L)}}
%\eduardo{Tree and PM forces combine.}
To combine the forces provided by the {\pmesh} and Tree codes, the gravitational potential is split into the sum of two fields: 
%\eduardo{Cite gadget2 for this equation.}
\begin{equation}    
    \Phi = \Philong + \Phishort\, ,
    \label{eq:phi_long_short}
\end{equation}
where $\Philong$ represents long-range modes from the {\pmesh}, and $\Phishort$ represents short-range modes from the Tree.
Written in Fourier space (tilde on top of symbols denotes a Fourier transform), the Poisson equation reads:
\begin{equation}
\tilde{\Phi}_k 
= -\frac{4\pi}{k^2} \tilde\rho_k\, ,
\end{equation}
where $\rho$ denotes the mass density. We can split the density as a sum of short-range and long-range terms, using Gaussian filters:
\begin{equation}
\tilde{\Phi}_k = -\frac{4\pi}{k^2} \tilde\rho_k \left( 1 - \exp(-k^2 r_a^2) \right)
   -\frac{4\pi}{k^2} \tilde\rho_k \exp(-k^2 r_a^2)\, .
\end{equation}
The scale $r_a$ is the one at which we split long- and short-range modes. 
We can obtain $\Phishort$ by solving the modified Poisson equation for short modes:
\begin{equation}
\tildePhishort_k = 
    -\frac{4\pi}{k^2} \tilde\rho_k \left( 1 - \exp(-k^2 r_a^2) \right)\, ,
    \label{eq:poisson_short}
\end{equation}
and $\Philong$ by solving the modified Poisson equation for long modes
\begin{equation}
\tildePhilong_k = 
    -\frac{4\pi}{k^2} \tilde\rho_k \exp(-k^2 r_a^2)\, .
    \label{eq:poisson_long}
\end{equation}

The long-mode Poisson equation (\ref{eq:poisson_long}) is solved by the {\pmesh} in Fourier space, so the convolution with the kernel is a simple multiplication.
The Tree on the other hand works in real space, hence equation (\ref{eq:poisson_short}) has to be transformed; this can be done analytically, yielding:
\begin{equation}
    \Phishort(\vec x)
    = -G \sum_{i} \frac{m_i}{|\vec x - \vec r_i|} 
        \erfc\left( \frac{|\vec x - \vec r_i|}{2r_a^2} \right)\, .
\end{equation}

%\eduardo{The \pmesh\ in \gadget\ takes into account the sampling and interpolation losses of power at short wavelengths, by convolving the gravitational potential with CIC correction.}
% \eduardo{Mention the Hamiltonian nature of the timesteps, ie. position and
% momentum describe the state of particles.}
% \eduardo{Domain decomposition.}
% \eduardo{Is is relevant to say that timesteps are adaptive?}

\subsection{\grgadget}
%\eduardo{We are developing a gevolution library. Why?}

\subsubsection{{\libgevolution} library}
\label{sec:libgevolution}

In order to have a relativistic {\pmesh} code working in {\gadget}, we developed a library that implements both the Newtonian and the relativistic {\pmesh} algorithms of the monolithic {\gevolution} code. This was done by forking the {\gevolution} github repository into {\libgevolution}, a library that is publicly available on github\footnote{\url{https://github.com/GrGadget/gevolution-1.2}} under MIT license.

%We are working on the developement of a C++ library that implements the concepts and tools associated to a relativistic \pmesh\ based on Fourier methods.
%The design goal of this library is to port \gevolution\ capabilities into other C/C++ codes such as \gadget.
%The user interface will consist of template functions and classes so that the library can be used with little constraints on the data model.

%\eduardo{Why is this work unique?}

% \eduardo{This section describes the work done to make this possible,
% implementation details relevant to \gevolution\ developers. First user
% guide.}
%\eduardo{How would we like the gevolution library to be structured?}
The rationale behind the development of \libgevolution\ is to encapsulate \gevolution's resources and methods into abstract objects. This yields several benefits.
Firstly, {\gevolution} maintenance is eased by the logical modularization of the code, i.e. instead of a monolitic code with a unique workflow we can divide {\gevolution} into components (C++ classes and/or namespaces) with well defined purposes.
Secondly, we are allowed to re-use {\gevolution} components within other applications, such as we do within {\gadget} in the present paper.

\begin{figure}
    \centering
    \begin{tikzpicture}[node distance=2cm]
        \node (pm) [class] {\texttt{particle\_mesh}};
        \node (newton) [class, below right of=pm]{\texttt{relativistic\_pm}};
        \node (gr) [class, below left of=pm]{\texttt{newtonian\_pm}};
        \draw [arrow] (gr) -- (pm);
        \draw [arrow] (newton) -- (pm);
    \end{tikzpicture}
    \caption{\pmesh\ class hierarchy in \libgevolution.}
    \label{fig:pm_hierarchy}
\end{figure}
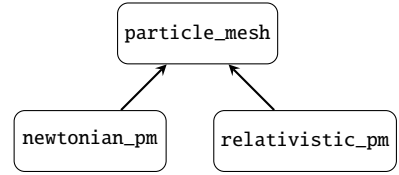

We give here an overview of the library; the precise signature of all the defined functions, methods and data structures is described in the technical documentation of the code.
{\libgevolution} is based on three cornerstones: (i) a particle container implemented through the class \verb|Particles_gevolution|;
(ii) a \pmesh\ data structure named \verb|particle_mesh|, templated on the particle container type, that can be used either as a \verb|relativistic_pm| or a \verb|newtonian_pm|;
(iii) an executable application that uses the previous components to produce N-body simulations as the original code does.
%\eduardo{The PM concept.}
\verb|particle_mesh| has to be understood as a container that is aware of the parallelization of the tasks and distribution of memory; it holds the gravitational fields 
%(the metric perturbation components in the case of relativistic \pmesh\ or the single gravitational potential in the case of Newtonian gravity) 
and it allows the user to compute the forces acting on the simulation particles.
The user interface declared in \verb|particle_mesh| consists of the following functions:
\begin{itemize}
    \item \texttt{sample(...)}, that builds the sources (density field or energy-momentum tensor) by sampling particle properties in the mesh;
    \item \texttt{compute\_potential(...)}, that solves Poisson equations to compute the potential fields;
    \item \texttt{compute\_forces(...)}, that computes the forces acting on particles.
\end{itemize}

\verb|particle_mesh| is specialized to solve the Newtonian problem or the General Relativistic problem using class inheritance;
Figure \ref{fig:pm_hierarchy} illustrates the class hierarchy of \libgevolution's \verb|particle_mesh|.
The expert user will be able to specialize \verb|particle_mesh| to his/her own needs, for example by deriving a {\pmesh} that solves a modified gravity problem.

% \eduardo{The newtonian PM.}
\verb|newtonian_pm| is the specialization of \verb|particle_mesh| that contains a real \verb|LATfield2::Field| scalar field %, represented as a real 
$\Phi_{\text{Newton}}$ and its complex \verb|LATField2::Field| Fourier transform $\tilde\Phi_{\text{Newton}}$,  plus a \verb|LATField2::PlanFFT| that connects 
$\Phi_{\text{Newton}}$ with $\tilde\Phi_{\text{Newton}}$ through discrete Fourier transform.
% \eduardo{The GR PM.}
\verb|relativistic_pm| is the specialization of \verb|particle_mesh| that contains the above quoted degrees of freedom of the perturbed \FLRW\ metric,
$\Phi$, $B_i$ and $\chi$. These are represented as real \verb|LATfield2::Field|, with complex \verb|LATField2::Field| counterparts to represent their Fourier transforms 
and a \verb|LATField2::PlanFFT| for each field.

As a first testing phase, we run {\libgevolution}, called with a simple wrapper, and the native {\gevolution} code, applying them to the same set of initial conditions, checking that the results were identical both in the Newtonian and relativistic cases. Then we stripped down {\gadget} by switching off the Tree code, and compared its results to the Newtonian results of {\libgevolution}. It is necessary that this comparison gives nearly identical results if we want {\libgevolution} to substitute the native {\pmesh} code of {\gadget} without loss of accuracy. To achieve a satisfactory match of the two PM codes we had to change the {\gevolution} scheme in a few points.

We started from V1.2 of {\gevolution}, that implemented a first-order version of finite differences instead of the fourth-order scheme of {\gadget}. This resulted in a difference with {\gadget} run on the same initial conditions, and in a percent-level offset of the matter power spectrum on large scales at low redshift. We upgraded the computation of spatial derivatives to fourth order, in parallel with the {\gevolution} developers that had noticed the same problem; our implementation is equivalent the most recent issue of {\gevolution} \citep[used, e.g., in][]{Adamek2022}.
The upgrade is the following: let's consider the gravitational potential along one direction of the mesh, and let's call its values $\Phi_i$, where the index $i$ denotes its position along that direction. 
Its first derivative is computed with finite differences at the first order as:
\begin{equation}
    \frac{\partial \Phi_i}{\partial x}
    = \frac{\Phi_{i+1} - \Phi_i}{h} + \mathcal{O}(h),
    \label{eq:fd1}
\end{equation}
where $h$ is the size of the mesh cell. Fourth-order Taylor expansion gives: 
\begin{equation}
    \frac{\partial \Phi_i}{\partial x}
    = 8 \frac{\Phi_{i+1} - \Phi_{i-1}}{12 h} 
    - \frac{\Phi_{i+2} - \Phi_{i-2}}{12h} 
    + \mathcal{O}(h^4)\, .
    \label{eq:fd4}
\end{equation}
This has a smaller error of order $\mathcal{O}(h^4)$, so it achieves higher precision than \eqref{eq:fd1} with the little cost of knowing the potential value at the second-nearest cell, that implies a negligible communication overhead.

% \eduardo{Some improvements to the code. CIC corrections and Poisson solver.}
Another improvement with respect to V1.2 of {\gevolution}, that follows an
implementation of {\gadget}, was the application of correcting filters to the
density in Fourier space to compensate for cloud-in-cell (CIC) interpolation.
Indeed, as discussed e.g. in \citet{Springel2005} or \citet{sefusatti2016}, CIC interpolation at some finite order leads to some loss of power that can be compensated for in Fourier space using suitable kernels. This was applied both to the computation of the density and to the computation of energy-momentum tensor components in the relativistic case.

Lastly, to make the Newtonian {\pmesh} scheme equivalent to that of {\gadget} we changed the form of the discrete Laplacian operator in the Poisson equation solver from its original form
\begin{equation}
    \nabla^2 \to 
    - \frac{4 N^2}{L^2}
    \Big( 
         \sin^2 \frac{\pi k_x}{N}  
        +\sin^2 \frac{\pi k_y}{N}  
        +\sin^2 \frac{\pi k_z}{N}  
        \Big)\, ,
\end{equation}
described in \cite{Adamek2016}, equation (C.5), to the form used in {\gadget}:
\begin{equation}
    \nabla^2 \to 
    - \frac{4\pi^2}{L^2}
    \Big( 
        k_x^2 + k_y^2 + k_z^2        
    \Big).
\end{equation}

\subsubsection{Calling {\libgevolution} from {\gadget}}
% \eduardo{This section describes how can we use \gevolution\ library from \gadget,
% implemetation details relevant to \gadget\ and \gevolution\ developers.}
% \eduardo{In order to use \gevolution\ within \gadget:
% \begin{itemize}
%     \item there must be data compatibility, or a communication channel,
%     \item unit conversion scheme
%     \item parallelization strategies,
%     \item modify \gadget\ drift method,
%     \item rethink the coupling of Tree+\pmesh.
%     \item we helped ourselves with the C++17 standard features and boost
%     libraries.
%     \item how is the time step decided?
%     \item time integration scheme.
% \end{itemize}}

%\eduardo{How do we use gevolution's PM into Gadget4? A new PM class.}
%In the previous subsection we have described the modularization of 
%\libgevolution\ that permits its use in higher levels applications such as
%\gadget. 

\begin{figure*}
    \centering\includegraphics[width=.8\textwidth]{%
        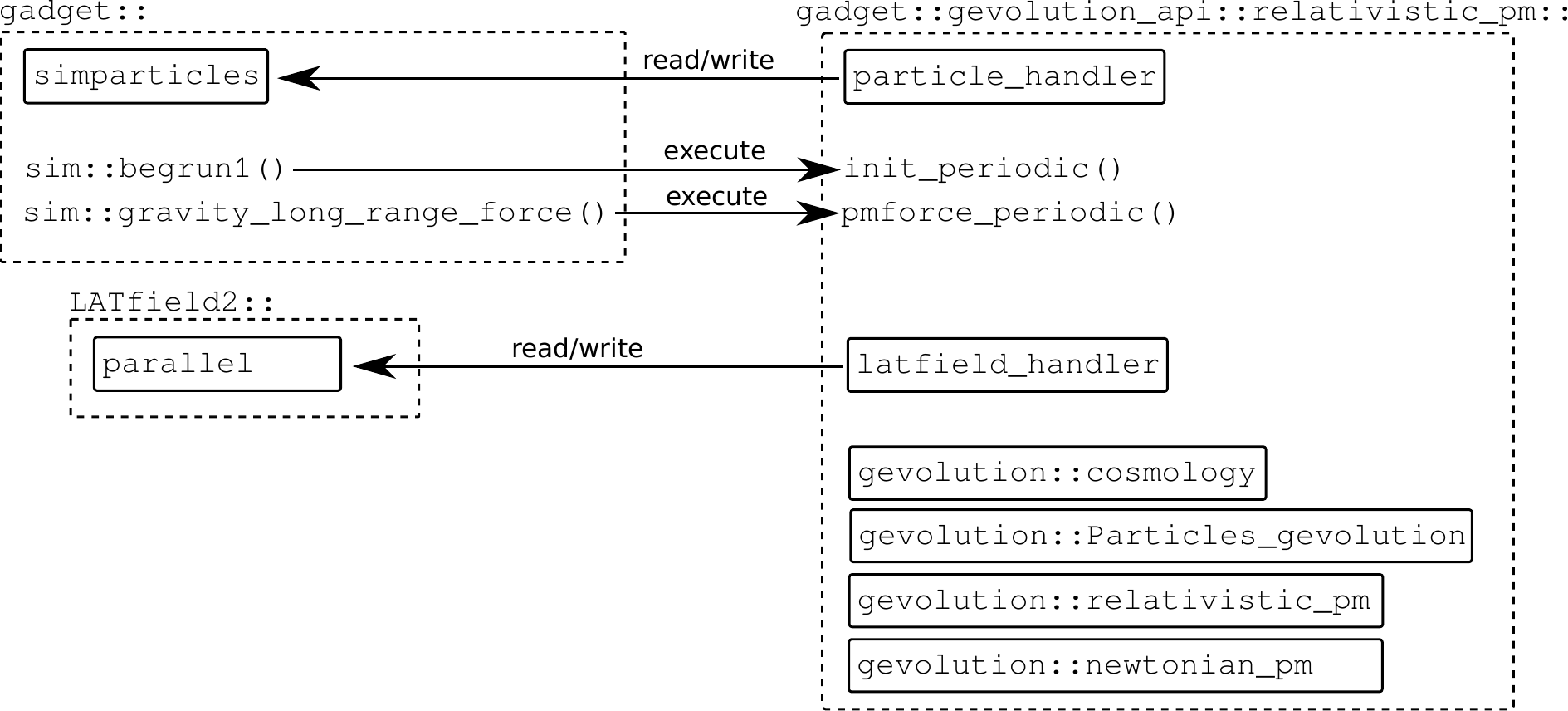}%
    \caption{Diagram of resource ownership and relations for \libgevolution\
    integrated into \gadget's workflow. 
    Each solid box represent a memory resource (an instantiation of a variable
    type) while the dashed boxes indicate ownership.
    The newly developed code, represented in
    the right part of the diagram denoted with the namespace
    \texttt{gadget::gevolution\_api}, consists in a class named
    \texttt{relativistic\_pm} that owns 
    a \texttt{particle\_handler} object that reads and writes directly into 
    \texttt{gadget::simparticles}, a
    \texttt{latfield\_handler} that takes care of setting up and inspect the
    state of \texttt{LATfield2::parallel},
    and some types defined in \libgevolution, that {\bf are} defined in
    \texttt{gevolution} namespace, like \texttt{cosmology},
    \texttt{Particles\_gevolution} and \texttt{relativistic\_pm}.
    The methods \texttt{sim::begrun1()} and
    \texttt{sim::gravity\_long\_range\_force()}
    in \texttt{gadget::} interact with the \texttt{relativistic\_pm} through
    their interface \texttt{init\_periodic()} and \texttt{pmforce\_periodic()}.}
    \label{fig:gevapi}
\end{figure*}

The implementation of {\libgevolution} in {\gadget} was performed as follows.
We created a new {\pmesh} class with a similar interface as the original one in \gadget, so that it is initialized and executed with the same functions as {\gadget}, i.e. \texttt{init\_periodic()}
and \texttt{pmforce\_periodic()}.
A new class \texttt{relativistic\_pm} was implemented within an
\texttt{gadget::gevolution\_api} namespace, avoiding to use the wider \texttt{gadget} namespace to make a clear distinction of purpose between the original {\gadget} code and our additional features.
% \eduardo{Implementation details of the PM. Data model, calling gevolution
% methods, domain decomposition.}
This \texttt{relativistic\_pm} class %in \texttt{gadget::gevolution\_api} namespace
acts much like a mediator taking information in and out of
\texttt{gadget} simulation particles, processing the correct units conversion
and calling the methods on \texttt{gevolution} namespace.
Figure \ref{fig:gevapi} shows a diagram that summarizes the contents of this \pmesh\ class, its relation with \gadget's resources and the entry points for \texttt{gevolution}'s api.

\texttt{relativistic\_pm} consists of:
\begin{itemize}
    \item A variable of type \verb|simparticle_handler| that acts as a wrapper for providing particle information
    %hides the
    %complications of reading particles' information (mass, position, momentum)
    from \gadget's \texttt{simparticles} global variable
    and writing back the data produced by \texttt{gevolution}'s \pmesh.
    \item A variable of type \verb|latfield_handler| that takes care of
    correctly initializing \latfield\ global state. Indeed, while {\gadget} can run
    with any number of MPI processes, \latfield\ has limitations that depend on
    the number of grid points in the \pmesh.
    \verb|latfield_handler| also takes care of creating
    a sub-communicator from \gadget's MPI global communicator that satisfies the
    constraints set by \latfield.
    \item A variable of type \texttt{gevolution::cosmology} that contains the
    parameters for the background evolution.
    \item A container of type \texttt{gevolution::Particles\_gevolution} that
    holds particle information, stored according to their location on the \pmesh\ grid.
    \item Variables of type 
    \texttt{gevolution::relativistic\_pm} 
    and
    \texttt{gevolution::newtonian\_pm} 
    that perform the actual \pmesh\ computations, i.e. construct the sources,
    either density or the components of the energy-momentum tensor, 
    compute the gravitational potential or the metric perturbation fields and
    the forces that act upon the particles.
    \item The methods \verb|pm_init_periodic| and
    \verb|pmforce_periodic|, for initialization and execution of the {\pmesh}, respectively.
\end{itemize}

\subsubsection{Kick and drift operators}

% \eduardo{State variables of the particles. Definition of force.}
In order to keep the Hamiltonian character of the equations of motion in
\gadget, we have to describe the state of each particle through its position and momentum, not velocity. Following a leap-frog scheme, the momentum should be updated with a \emph{kick} operation using the full relativistic Eqs.~\eqref{eq:hamilton_eq1} and \eqref{eq:hamilton_eq2}. However, velocities in {\gadget} are to be interpreted as momenta (per unit mass) of non-relativistic particles in the Newtonian limit. Then we redefine the {\gadget} \emph{kick} and \emph{drift} operators assuming non-relativistic matter, $p\ll m ca$, and further neglecting the very small contribution coming from $\chi$:

\begin{equation}
    \begin{split}
    \frac{dx^i}{d\tau}
    = 
    & \frac{p^i}{ma}\left( 1 + 3\Phi \right)
    + c B^i\, ,
    \end{split}
    \label{eq:hamilton_eq3}
\end{equation}

\begin{equation}
    \begin{split}
    \frac{dp_i}{d\tau}=
    &
       -c p^n B_{n|i}
       - \Phi_{,i} m c^2 a\, .
    \end{split}
    \label{eq:hamilton_eq4}
\end{equation}

%and the position is updated with a \emph{drift} operation using equation \eqref{eq:hamilton_eq3}. 
\noindent
The right hand side of \eqref{eq:hamilton_eq4}  is what we call \emph{force}.
%\eduardo{Mention the fix to the drift function.}

\subsubsection{Adding long-range and short-range forces}

% \eduardo{Combine PM and Tree forces.}
To combine the forces computed with the relativistic \pmesh\ and \gadget's Newtonian Tree
we have extended the idea of the \treepm\ coupling.
From equation \eqref{eq:phi_long_short} one obtains that the force acting on
a particle in a \treepm\ scheme consists of two terms:
\begin{equation}
    \vec F = S_{r_a} [\vec F^{\mathrm{Tree}}_{\mathrm{Newton}}] 
        + L_{r_a}[\vec F^{\mathrm{PM}}_{\mathrm{Newton}}].
    \label{eq:force1}
\end{equation}
The first term is the force computed using the Tree on which an exponential
high-pass filter $S_{r_a}$ is applied, leaving short-wavelength modes. The second term corresponds to the \pmesh\ force on which the complementary low-pass filter $L_{r_a}$ is applied to leave long-wavelength modes.
The symbols $S_a$ and $L_a$ formally denote these linear operators:

\begin{equation}
    S_{r_a}[ f ](\vec r)  
    = \frac{1}{N} \sum_{\vec k} \tilde f_{\vec k} (1 - \exp(-k^2 {r_a}^2)) \exp( - i
    \vec k \cdot \vec r)\, ,
\end{equation}
and
\begin{equation}
    L_{r_a}[ f ](\vec r)  
    = \frac{1}{N} \sum_{\vec k} \tilde f_{\vec k} \exp(-k^2 {r_a}^2) \exp( - i \vec
    k \cdot \vec r)\, .
\end{equation}

\noindent
The \emph{grid smoothing scale} $r_a$ scales with the {\pmesh} mesh size, and
its value is optimized in {\gadget}, in a way that will be tested below, to
minimize the impact of the two different treatments of the gravitational force.

In order to account for the relativistic dynamics while preserving the match
between tree and PM contributions that is valid in the Newtonian case, we choose
the following strategy: {\gadget} calls both {\tt newtonian\_pm} and {\tt
relativistic\_pm}, the Newtonian value of the force is added to the Tree force
as in a standard Newtonian simulation, while the difference between the
Newtonian and the relativistic forces is added on top as a correction, but
filtered on a different scale $r_b$, that we call \emph{gr-smoothing scale}. 
Eq.~\eqref{eq:force1} then becomes:
\begin{multline}
    \vec F = 
    S_{r_a} [\vec F^{\mathrm{Tree}}_{\mathrm{Newton}}] + 
    L_{r_a}[\vec F^{\mathrm{PM}}_{\mathrm{Newton}}]
    % \\
     %- L_{r_b}[]
     + L_{r_b}[\vec F^{\mathrm{PM}}_{\mathrm{GR}}-\vec F^{\mathrm{PM}}_{\mathrm{Newton}}]\, .
    \label{eq:force2}
\end{multline}
The case $r_a=r_b$ would correspond to simply adding the relativistic force to the Tree:
\begin{equation}
    \vec F = 
    S_{r_a} [\vec F^{\mathrm{Tree}}_{\mathrm{Newton}}]
     + L_{r_b}[\vec F^{\mathrm{PM}}_{\mathrm{GR}}]\, .
    \label{eq:force3}
\end{equation}
However, while the size of $r_a$, that regulates the match between Newtonian Tree and PM forces, is very well tested within {\gadget}, the optimal value of $r_b$ is to be found; we will show in the next Section that using $r_b$ larger than $r_a$ allows us to achieve percent accuracy at small scales.

%\eduardo{We are using a relativistic PM in Gadget.}
%With the \libgevolution\ at hand (even at an early stage of developement) we have exploited the similarities between the \pmesh\ in \gadget\ and \gevolution\ to produce a version of \gadget\ in which the Tree is complemented with a relativistic \pmesh. In this approach we are assuming that the equations for the relativistic gravity in the Poisson gauge will converge to those of the Newtonian gravity at distances smaller than the \pmesh\ resolution, that means the cell size must be much smaller than the Hubble radius. With this implementation we seek to produce high resolution and realistic cosmological simulations: on one hand there is the non-linear clustering that the Tree will be able to reproduce and on the other hand we will be able to look into relativistic effects on-the-fly at the scales permitted by the \pmesh. Because the matter inhomogeneities, represented here as particles, generate the perturbed metric we would expect that the non-linear features (high density regions and vorticity) enhanced by the Tree will contribute to the strength of the short wavelength modes of these metric perturbations.
\section{Validation}

The {\grgadget} code has been validated by running it on a few realizations of initial conditions, listed in table~\ref{tab:list_configurations}.
These %initial conditions 
were generated with {\gadget}'s \texttt{ngenic} code
at $z=19$, starting from a linear power spectrum generated with 
\textsc{CAMB}\footnote{\url{https://camb.info/}} and with cosmological parameters consistent with Planck 2018 result \citep{planck2018}:
$\Omega_b h^2 = 0.0223$, $\Omega_c h^2 = 0.120$, $H_0 = 67.3\, \SI{}{km\,s^{-1}\,
Mpc^{-1}}$, $A_s = 2.097\times 10^{-9}$ and $n_s=0.965$. 

\begin{table}
    \centering
    \begin{tabular}{c|ccc}
        name & $N_p$ (particles) & $N$ (\PM\ grid points) & $L$ (box size) \\ 
        \hline
        \testsmall & $64^3$ & 64 & $1\,\SI{}{Gpc}/h$\\
        \testmed & $256^3$ & 256 & $1\,\SI{}{Gpc}/h$\\
        \testbig & $512^3$ & 512 & $500\,\SI{}{Mpc}/h$\\
        \hline
    \end{tabular}
    \caption{Cosmological simulation configurations used
    to validate {\grgadget}.}
    \label{tab:list_configurations}
\end{table}

% \eduardo{This section must convey the following messages:
% 0. motivation for performing the tests we have performed,
% 1. we have tested the code at various levels of scope,
% 2. each test has a concluding remark,
% 3. overall these results validate the code and our main hypothesis.}

\subsection[Gevolution and Gadget-4 original codes]{{\gevolution} and {\gadget} original codes}\label{sec:original_codes}
% In section \ref{sec:libgevolution} we have discussed some changes we have
% introduced in the back-end of \libgevolution's numerical algorithms that we
% deemed as improvements with respect to the original \gevolution\ code.
% The choice for a specific back-end in scientific computing is generally
% justified if it favours one desired feature over another. 
% For example if we are more interested in obtaining a faster code at the expenses
% of accuracy, or maybe the typical problem we are dealing with is better
% solved with a code that does not perform well in other domains.
% While designing \libgevolution\ we had in mind its incorporation in \gadget,
% hence we had chosen a back-end that could reproduce with accuracy the results
% that we could obtain with \gadget's original \pmesh. The reason why, 
% is simply to be able to have a common baseline in Newtonian simulations either by
% running \gadget\ original code or by running \grgadget\ in Newtonian mode.
% Here we will present some tests we have perfomed to justify and validate the
% changes in \libgevolution's back-end.

As already discussed in Section~\ref{sec:libgevolution}, the {\tt newtonian\_pm} implementation in V1.2 of {\gevolution} computes the Newtonian forces differently from those obtained with \gadget's \pmesh. Before implementing {\libgevolution} as the {\pmesh} engine of {\gadget}, we need to make the two algorithms work in the same way. 

%\eduardo{Explain how is this test performed so that results can be reproduced by
%others.}

\begin{figure*}
    \centering
    \includegraphics[width=.48\textwidth]{%
    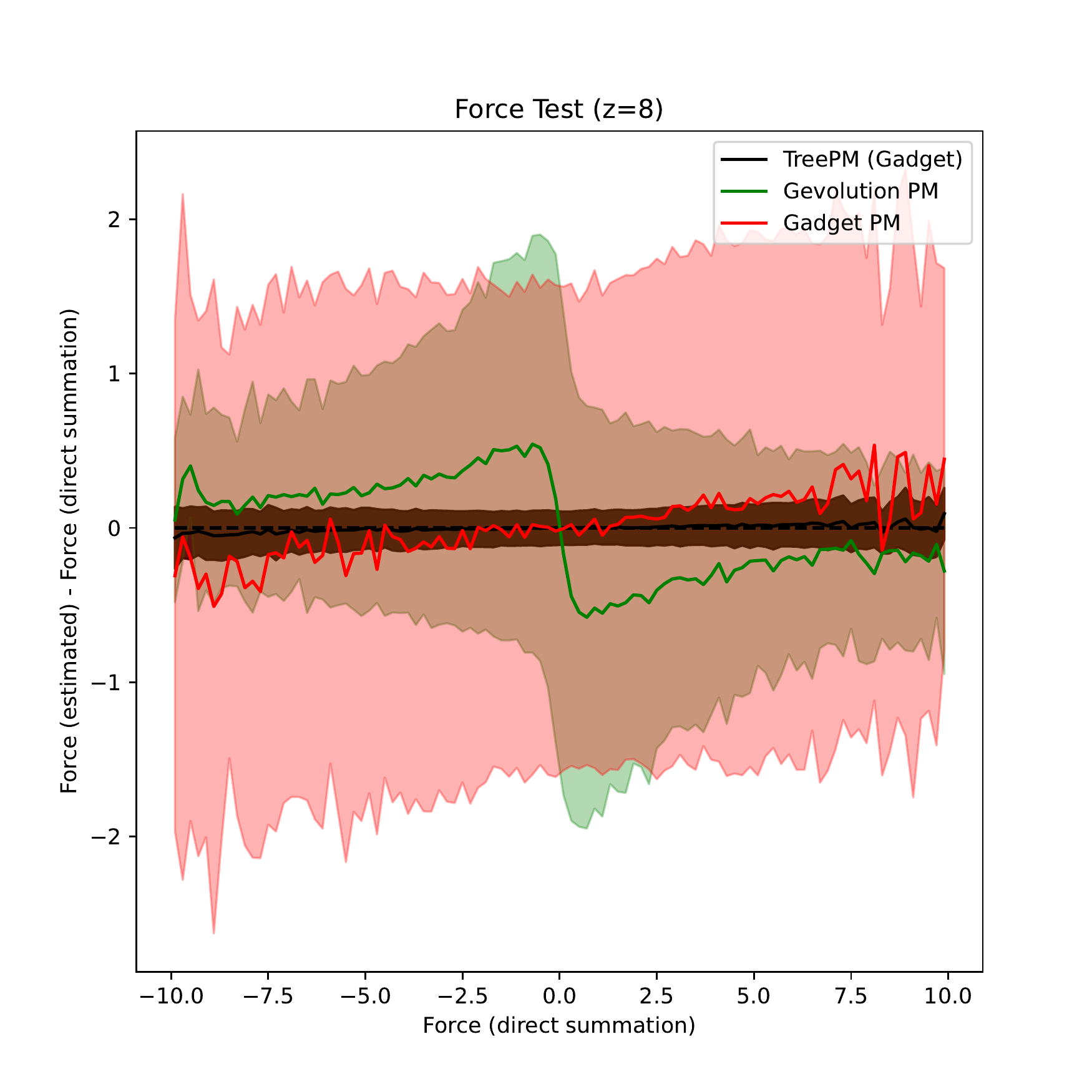}%
    \includegraphics[width=.48\textwidth]{%
    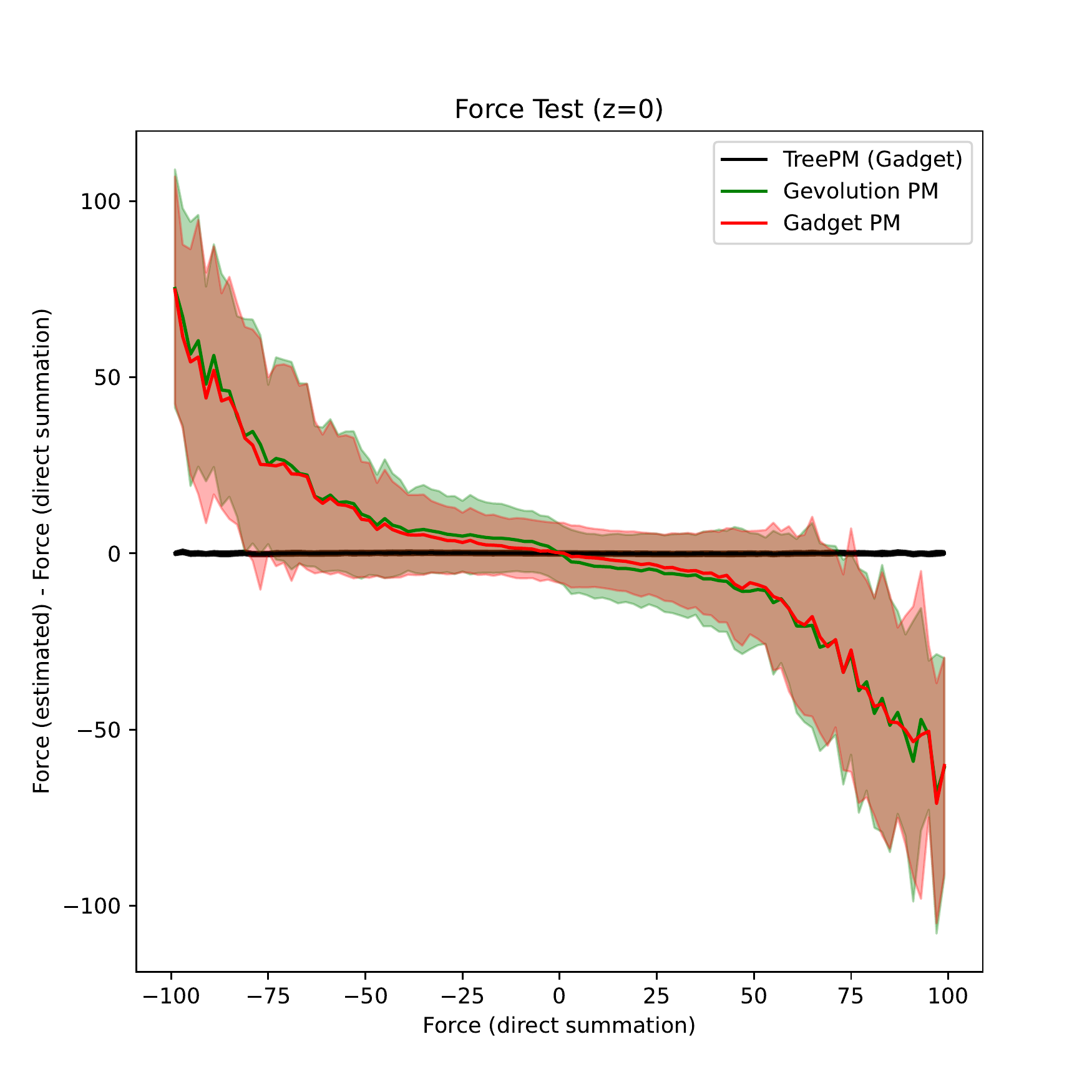}
    %\caption{Force dispersion with respect to direct summation
    %for a typical cosmological distribution of
    %dark matter particles, at $z=8$. The lines represent the mean value of the
    %force dispersion on different force bins
    %and the
    %shaded regions indicate the variance.}
    %  \item $z=8$
    %  \item $N = 64^3$ particles
    %  \item $M = 64^3$ grid size
    %  \item $L = \SI{1}{Gpc}/h$
%\end{figure}
%\begin{figure}
    \caption{Difference of gravity force with respect to the true PP value, binned according to the true force,
    for {\testsmall} initial conditions, at $z=8$ (left panel) and  $z=0$ (right panel). Lines represent the mean value of force difference in the bins, with colours explained in the legend; the shaded regions give the standard deviation of the corresponding force difference. }
     \label{fig:forcetest}
    %  \item $z=8$
    %  \item $N = 64^3$ particles
    %  \item $M = 64^3$ grid size
    %  \item $L = \SI{1}{Gpc}/h$
\end{figure*}

To this aim, we have run a set of simulations with the configuration \testsmall\
(described in table~\ref{tab:list_configurations}) with a small number of particles
$N_p=64^3$ to be able to compute forces using a straightforward particle-particle (PP) scheme, that can be taken as the true force that we are trying to approximate.
The same initial conditions at $z=19$ have been fed to both {\gadget} (with Tree either on or switched off to have a pure PM run) and {\gevolution} (in Newtonian mode) codes.
At later times, $z=8$ and $z=0$, we have written snapshots of the forces that the simulation particles experience, separating the {\pmesh} and the
{\treepm} components; we have then compared those to the true Newtonian
force computed with the PP scheme.
%\eduardo{Interpret the results.}
The data we have obtained are summarized in the plots shown in figure
\ref{fig:forcetest}.
We have binned particles according to the value of the true force, then for each bin we have computed the mean (colored lines) and standard deviation (shaded regions) of the difference between the force computed with approximate methods ({\pmesh} or {\treepm}) and the true value.
Forces are given in {\gadget}'s default units, which is actually acceleration,
measured in units of
$10 H_0\ \SI{}{km/s} = h\, \SI{}{km^2\, s^{-2}\, kpc^{-1}}$.
The green line shows the {\pmesh} result using the original
{\gevolution} code (the true force is anyway computed with {\gadget} and matched particle by particle)
while the red line is obtained from a pure {\pmesh} using {\gadget}'s original
code.
The black line gives the {\treepm} method precision, obtained using {\gadget}.

Looking at the red and green lines (and their shaded areas) we find two known results. Firstly, the \treepm\ method produces far less bias and dispersion when estimating forces; for instance, in the left panel of Fig.~\ref{fig:forcetest} the error is of the order\footnote{This quantification is in code units, we can take this value as a reference for a high accuracy gravity solver.}
of $0.1\, h\,\SI{}{km^2\, s^{-2}\, kpc^{-1}}$, while in the right panel it is larger but barely visible when compared with the other curves. Secondly, while the {\pmesh} force has low bias but a much larger variance than the {\treepm} one at high redshift, at low redshift, i.e. at higher level of non-linearity, it underestimates the value of the Newtonian force as its magnitude increases.
This underestimation is due to the failure of {\pmesh} in resolving interaction at scales smaller than the grid resolution.

When comparing {\gevolution} {\pmesh} and true forces, we notice an \emph{S}-shaped feature in the plot, much more visible at high redshift. As anticipated in Section~\ref{sec:libgevolution}, this is mostly due to the first-order interpolation used to find the gradient of the potential in the code version that we tested.

%After changing the code as described in Section~\ref{sec:libgevolution}, the {\pmesh} force computed by {\libgevolution} called within {\gadget} gives results that are indistinguishable from those of the native {\pmesh} code.

%\eduardo{Testing the matter power spectrum.}
%\eduardo{Explain the test detail for reproducibility.}
%\eduardo{Interpret the results.}

In Fig.~\ref{fig:power_first} we show the matter 
power spectra\footnote{In this paper all particle power spectra were computed
using \texttt{PowerI4} code presented in \cite{sefusatti2016}. Unless otherwise
stated, all power spectra are computed up to the the Nyquist
frequency of the PM mesh.}
obtained at $z=0$ from a set of larger simulations with the configuration 
{\testmed} (see table~\ref{tab:list_configurations}).
The red solid line shows the result obtained with the original {\gadget} code with its {\treepm} method, while the red dotted line shows the results obtained by switching off the Tree so that the forces are computed using the {\pmesh} alone.
The green lines show results obtained with the latest {\em develop} version of  {\gevolution} that implements higher order schemes for finite differences; the dotted line gives results obtained with \texttt{GRADIENT\_ORDER=1} and is identical to the result obtained with V1.2 of {\gevolution}, the green solid line uses \texttt{GRADIENT\_ORDER=2}, that corresponds to a second-order scheme.
These power spectra show that the matter distribution in {\gevolution} using first-order gradients loses power in what seems to be a uniform trend for large-scale modes. This is a behaviour which is not inherent to the {\pmesh} nature
of the code, since that type of numerical approximation should predict very well
the linear evolution at large scales; indeed, the higher-order scheme recovers power on large scales to sub-percent accuracy.
Conversely,
{\gadget}'s {\pmesh} and {\treepm} agree very well at wavenumbers below $k\sim0.1h/{\rm Mpc}$ scale,

The higher-order differentiation worsens the loss of power of {\gevolution}
for high values of $k$, that is not present in {\gadget}.
This can be explained as a consequence of the particle-to-mesh sampling 
and mesh-to-particle interpolation described in section~\ref{sec:libgevolution}.
As discussed there,
{\gadget}'s {\pmesh} corrects for these effects, resulting in a power spectrum
that degrades only at very high values of $k$ as we approach the Nyquist frequency, while producing a $\sim2$ percent overcorrection at $k\sim0.4\ h/{\rm Mpc}$.

After implementing the higher-order differentiation scheme, the correction for the loss of power discussed above and the change in the discrete Laplacian operator (Section~\ref{sec:libgevolution}), the results of native {\gadget} and {\libgevolution} PMs become indistinguishable.

{  
  \def\path{images/gad-gev-first-power}%
  %\only<1>{\centering\includegraphics[height=\textheight]{\path/pw-002.pdf}}%
  %\only<2>{\centering\includegraphics[height=\textheight]{\path/pw-008.pdf}}%
  %\only<1>{\centering\includegraphics[height=\textheight]{\path/pw-002.pdf}}%
  %\only<2>{\centering\includegraphics[height=\textheight]{\path/pw-008.pdf}}%
  \begin{figure}
    \centering\includegraphics[width=\columnwidth]{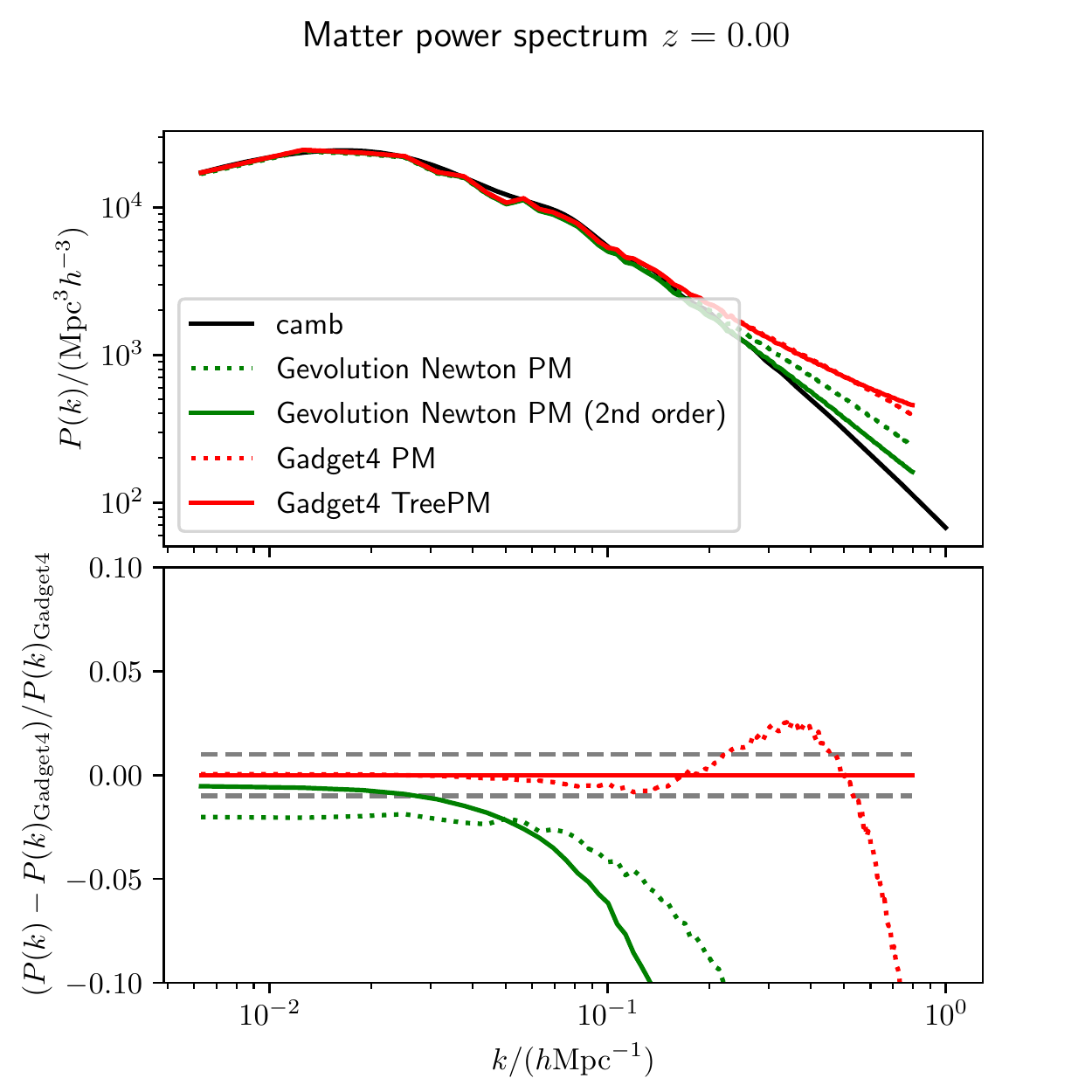}
    \caption{Matter power spectrum of {\testmed} cosmological simulations. The lower panel shows residuals with respect to \gadget's original code (in red), used as baseline. The black line shows the linear power spectrum obtained with CAMB.
    Red lines show results obtained with {\gadget}, with the Tree part on (solid line) or switched off (dotted line). Green lines show results obtained with {\gevolution} in Newtonian configuration, with finite differences at first order (dotted line) or second order (solid line).}
    \label{fig:power_first}
  \end{figure}
}

\subsection{Newtonian forces}\label{sec:test_newton}
%\eduardo{Newtonian force-test with gevolution library. Description of the test.}
% With \libgevolution\ we are offering a clean programatic way to execute
% \gevolution's code on a variety of user applications.
% We have, from the begining, exposed our intention to use \libgevolution\ to
% study General Relativiy in \gadget\ simulations.
% But as the tests shown in section \ref{sec:original_codes} are indicating that
% the numerical implementation of the \pmesh\ forces in \gevolution\ are somewhat
% different and producing measurable differences with respect to \gadget's \pmesh.
% That is why we have also worked to optimize \libgevolution\ in the sense of
% precision seeking to reproduce the Newtonian forces and power spectrum obtained
% with \gadget.
%In this section we are going to show the results of some tests we have performed to validate {\grgadget} code.

We have tested our implementation of the {\grgadget} code by running a standard test in {\gadget}:  we 
%The first test we are going to describe consists of re
create an N-body configuration in which there is a single massive particle in the entire simulation box, while other massless test particles are placed at different distances from the first. In this setting the exact value of the force on each particle is known,
%every particle experiences a 
%Newtonian force that is produced by the single influence of the first massive particle, 
hence one can compare the numerical results coming from the {\treepm}
algorithm to the analytical solution. 
%This is a standard test in
%{\gadget}, and we stress out that the results of this test performed with
%{\grgadget} matches the precision that was already reported in
%{\gadget} paper \citep{Springel2021}. 

\begin{figure}
    \centering\includegraphics[width=.45\textwidth]{%
        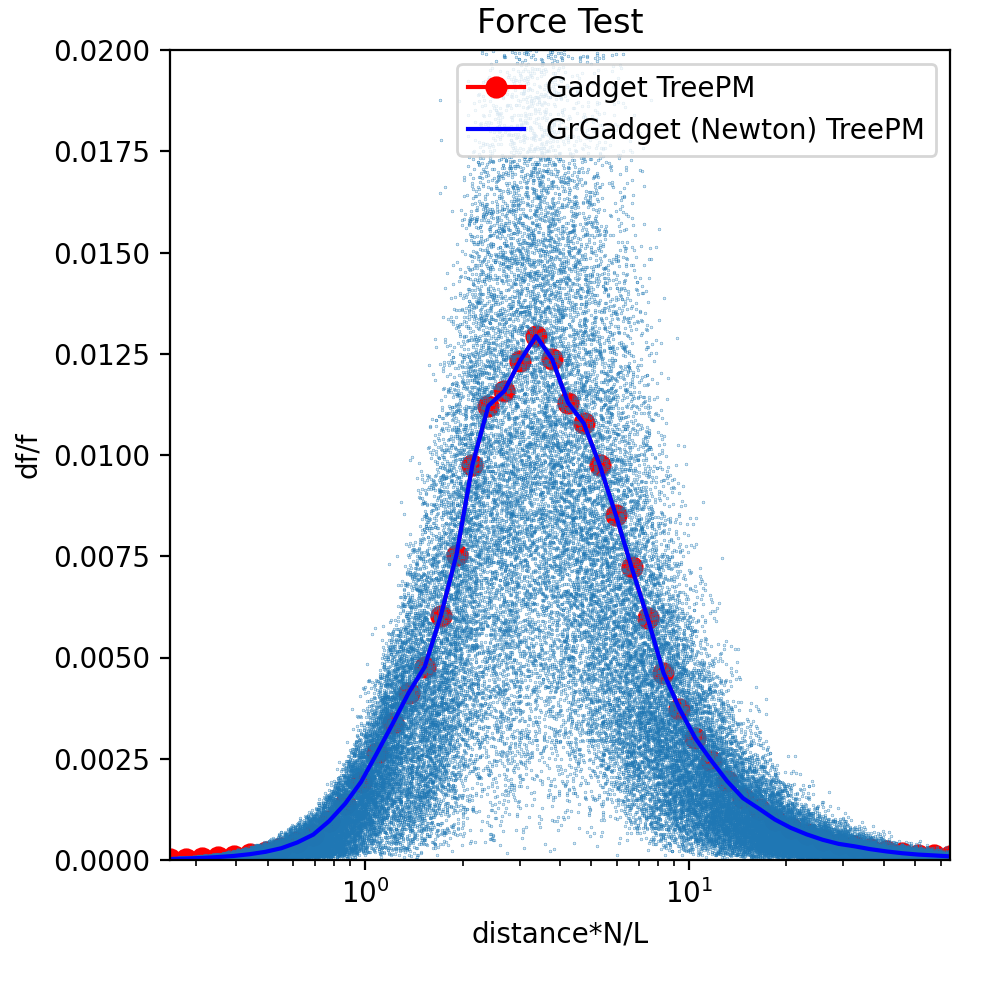}%
    \caption{Forces due to a point source:
    the points are test particles located at different distances (in units of
    the mesh resolution $L/N$) from the source
    and the lines represent the
    RMS of the difference between real and TreePM forces in different distance
    bins.
    The red line corresponds to \gadget\ original \treepm\
    while the blue line was obtained with {\grgadget} in 
    Newtonian mode. As for the the grid smoothing scale, the default value was used:
    $r_a = 1.25 L/N$.
    For this test we have used $N=256$ and $L=1\,\SI{}{Gpc}/h$.}
    \label{fig:dist_forcetest}
    % \item $z=0$
    % \item $N = 64^3$ particles
    % \item $M = 256^3$ grid size
    % \item $L = \SI{1}{Gpc}/h$
\end{figure}

%\eduardo{Explain the results.}
The results are shown in figure \ref{fig:dist_forcetest}, where
each dot represents a test particle.
The x-axis gives the distance to the massive particle that sources the gravitational field, in
units of the {\pmesh} resolution ($L/N$), while
the y-axis gives the corresponding absolute
value of the relative difference of the true and estimated forces acting on the test
particle. The red and blue lines correspond to the mean value of force residuals,  for particles binned into distance bins;
the red line denotes the statistics obtained from a simulation using {\gadget}'s
original {\treepm} implementation and the blue line was produced using
{\grgadget}, in this case with the Newtonian gravity engine.
% This test explores the precision of the two methods for computing
% forces at long and short distances and the way they are combined at the
% intermediate scales.

%\eduardo{Get to conclusions.}
This figure shows that the accuracy with which the {\treepm} code reproduces the gravitational force is at worst at percent level on scales of a few mesh cells,  corresponding to the scale where the PM and Tree contributions are matched, and gets very accurate in the limits where either the Tree (small scales) or the PM (large scales) dominates.

%From this figure we can conclude that %in both regimes:
%At short distances, right below the grid resolution
%where the Tree forces are dominant,
%and at long distances, from 10 times the grid resolution
%where the {\pmesh} forces are dominant, the estimated force by numerical methods
%is precise at the subpercent level and better as we approach the extremes.
%On the other hand, in the intermediate scales the combined Tree and {\pmesh}
%forces produce a numerical estimate that reaches on average
%a percent error in the worst distance bin.
%Finally we can also conclude both 
{\gadget}'s and {\grgadget}'s Newtonian {\pmesh}s show basically the same accuracy, even though their {\pmesh} implementations are very different. %Naturally, both implementations are conceptually using
%the same procedure to resolve the forces.

In Fig.~\ref{fig:power_final} we show the matter power spectra of a set of
{\testmed} simulations (see table \ref{tab:list_configurations}). 
In this case we are comparing the matter clustering of
{\grgadget}, in blue (with Newtonian forces for testing purposes), against
{\gadget}, in red. In agreement with the previous test of force differences,
we find that both codes produce the same matter power spectrum up 
to floating point errors. This is verified both in the case of simulations
computing forces using a pure {\pmesh} and in the case of {\treepm}.

%Even though the power spectrum does not give all information contained in the matter distribution,
%(for example non-gaussianities are not present) 
%this test gives a strong indication that further deviations of measured
%quantities will be caused by changes in the underlying physics (i.e. when moving from Newtonian to relativistic gravity, or by introducing modified gravity) and not because of significant systematic errors caused by numerical artifacts.
{  
  \def\path{images/gad-gev-latest-power}%
  %\only<1>{\centering\includegraphics[height=\textheight]{\path/pw-002.pdf}}%
  %\only<2>{\centering\includegraphics[height=\textheight]{\path/pw-008.pdf}}%
  \begin{figure}
    \centering\includegraphics[width=.45\textwidth]{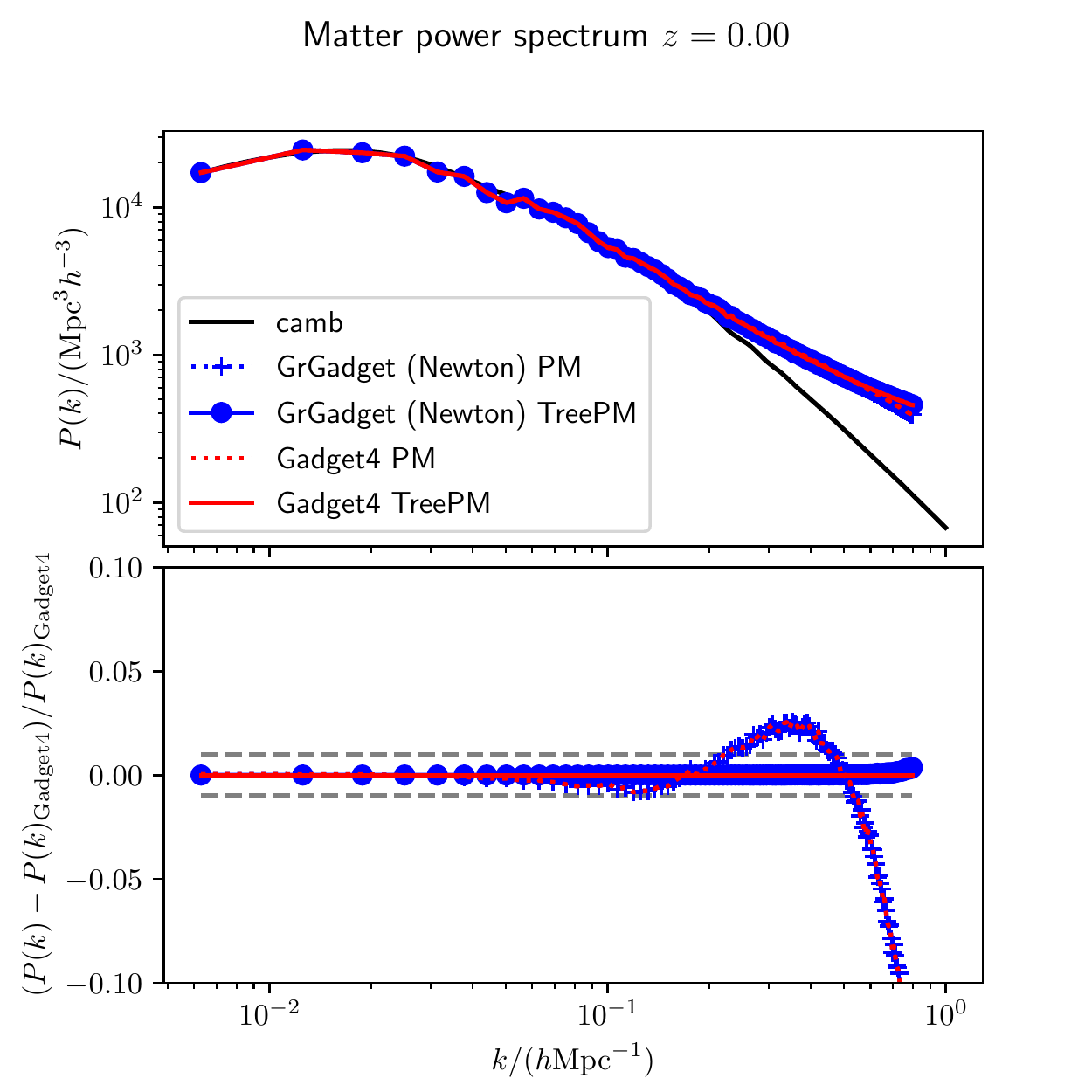}
    \caption{Matter power spectrum of four
    simulations starting from the same initial conditions \testbig: blue lines give results for
    \gadget\ original code, red lines give results for \grgadget. In both cases dotted lines refer to runs with \pmesh-only, solid lines refer to runs with full \treepm.}
    \label{fig:power_final}
  \end{figure}
}

\subsection{Relativistic simulations with {\grgadget}.}
% \eduardo{The matter power spectrum for the relativistic TreePM reproduces the
% correlations observed in \gevolution\ relativistic runs in the large scales and
% \gadget's at the non-linear scales. Clearly we don't know that would the
% relativist effects look like in the small scales unless we construct a finer
% mesh or modify the Tree forces.}

%{\grgadget} was conceived as an attempt to explore general relativistic effects
%in the high dynamic range of resolutions that a {\treepm} code can deliver.

We present here results obtained by running {\grgadget} with {\tt relativistic\_pm}, comparing them with the corresponding relativistic version of {\gevolution}. 
%general relativistic cosmological simulation of dark matter-only particles
We expect that the power spectrum of the matter density displays some
relativistic features at large scales due to terms preceded by $\daa$ in the
field equation (\ref{eq:einstein_perturbations_tt_gev}), while at
small scales results should be compatible with \gadget's Newtonian simulations.
However, the matter power spectrum shown here is not an observable quantity, so this comparison is just meant to give a first validation of the results. A more thorough comparison of observables reconstructed on the past light cone will be presented in a future paper.

Figure \ref{fig:power_gr_first} shows the matter power spectra for a series of
{\testmed} simulations (see table~\ref{tab:list_configurations}).
In this case {\gevolution} and {\grgadget} 
are run in GR mode. The parameter that regulates the scale of the
relativistic correction (Eqs.~\ref{eq:force2} and \ref{eq:force3}) is set to $r_b = 6\, L/N \approx 23 \SI{}{Mpc}/h$,
i.e. the relativistic corrections of the {\pmesh} method are smoothed at a
distances below $6$ grid cells.
The plot shows that
relativistic \pmesh-only simulations,
{\grgadget} (blue dotted line) and {\gevolution} (green lines)
are compatible on large scales ($k< 0.03 h/\SI{}{Mpc} $) up to a small percent-level difference that it is likely caused by the use of different orders for finite difference gradient; indeed, going from first- to second-order differences (from dotted to solid green line) the power spectrum gets nearer to {\grgadget}'s fourth-order one.
The plot also confirms that our combination of Tree and PM forces in the 
relativistic weak field limit with {\grgadget} (blue solid line)
reproduces the Newtonian non-linear features to sub-percent level at small scales, 
that is for $k>0.1 h/\SI{}{Mpc}$;
here {\gadget} (red solid line) is again our reference for the non-linear 
clustering. 

{  
  \def\path{images/gad-gev-gr-power}%
  \begin{figure}
    \centering\includegraphics[width=.45\textwidth]{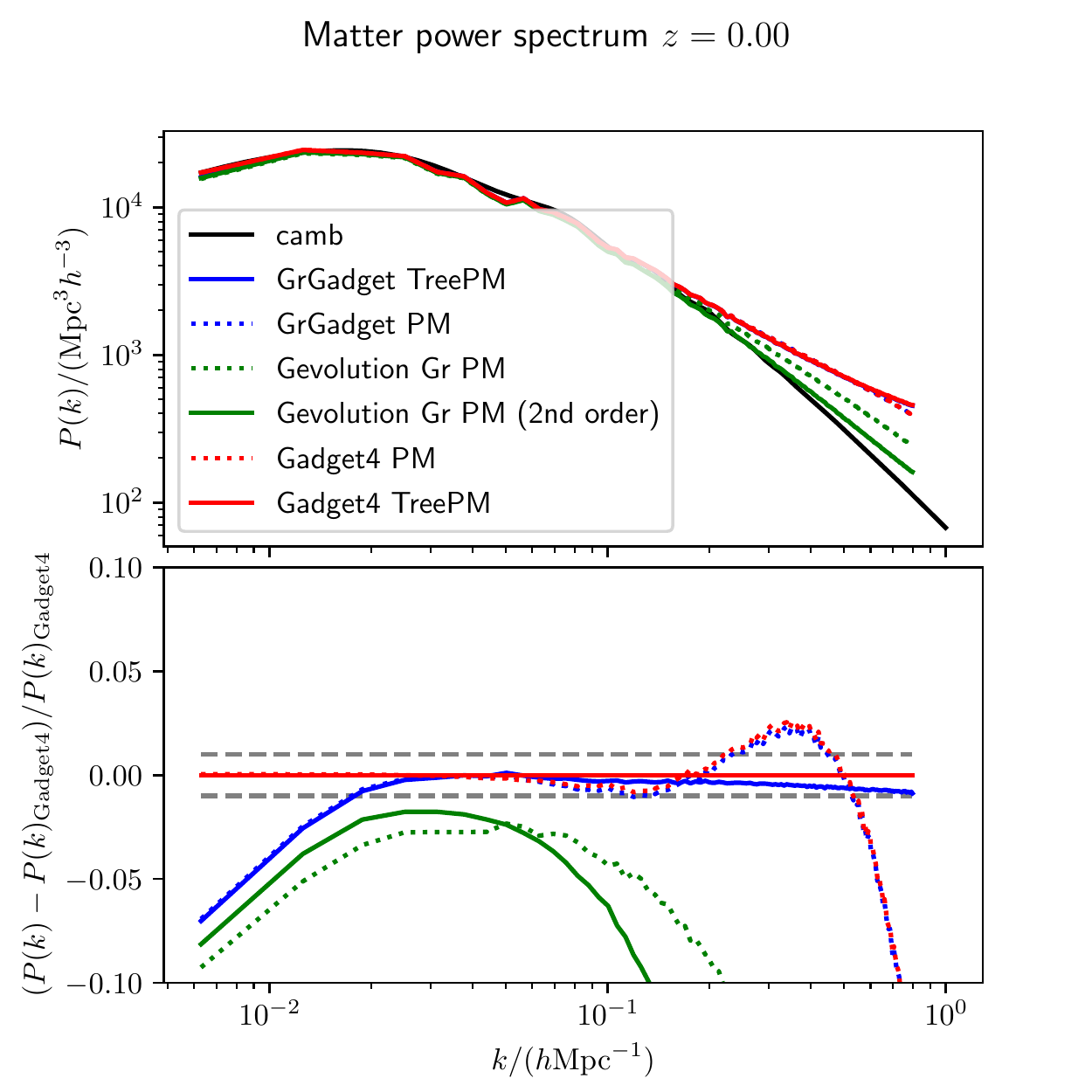}
    \caption{Matter power spectrum of
    \gadget, \gevolution\ and \grgadget\ runs, the last code being run in relativistic mode.
    The upper panel shows the absolute value and the lower panel the relative
    difference with respect to \gadget's \treepm.
    The black line gives the linear matter power spectrum; red and blue lines give {\gadget} and {\grgadget} results, with full TreePM forces (solid lines) or with the Tree switched off (dotted lines). Green lines give {\gevolution} results, dotted line referring to first-order finite differences (\texttt{GRADIENT\_ORDER=1}) and solid line referring to second-order calculation (\texttt{GRADIENT\_ORDER=2}).}
    \label{fig:power_gr_first}
    % gr-cut = 6.0
    % N = 256
    % L = 1000 Mpc/h
  \end{figure}
}

%with \gadget's Newtonian \pmesh\ (the red dotted line) at the small scales 
%for $k > 0.1 h/\SI{}{Mpc}$, this observation is coherent with the statement that
%for non-relativistic matter the Newtonian limit is recovered at distance well
%below the Hubble radius. Also it can be noticed that \gevolution\ loses power
%consistently at those scales, and we can attribute that effect mainly to the
%sampling and interpolation CIC filters, which are corrected in \gadget\ and
%\grgadget.
% The \pmesh\ method in \grgadget\ is compatible with \gevolution\ (the
%green line) in the
%large scales, for $k > 0.1 h/\SI{}{Mpc}$.
% The power spectrum corresponding to the \treepm\ in \grgadget\
%(represented with the contiguous blue line) is very well
%consistent with \gadget's original \treepm\ (the red contiguous line) 
%at the non-linear scales and it
%bends like \gevolution's GR feature does at large scales.

\begin{figure*}
    \centering
    \includegraphics[width=.49\textwidth]{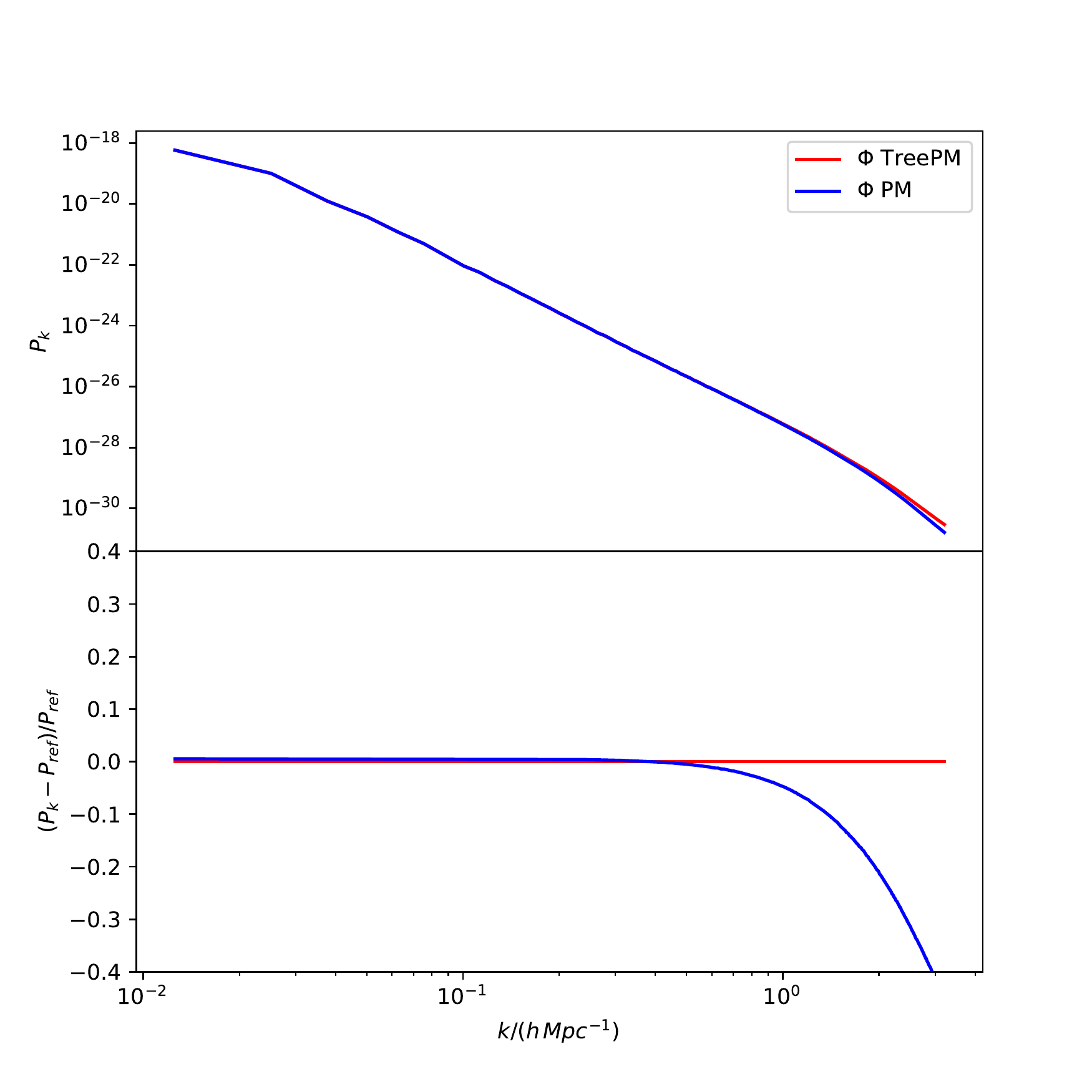}
    \includegraphics[width=.49\textwidth]{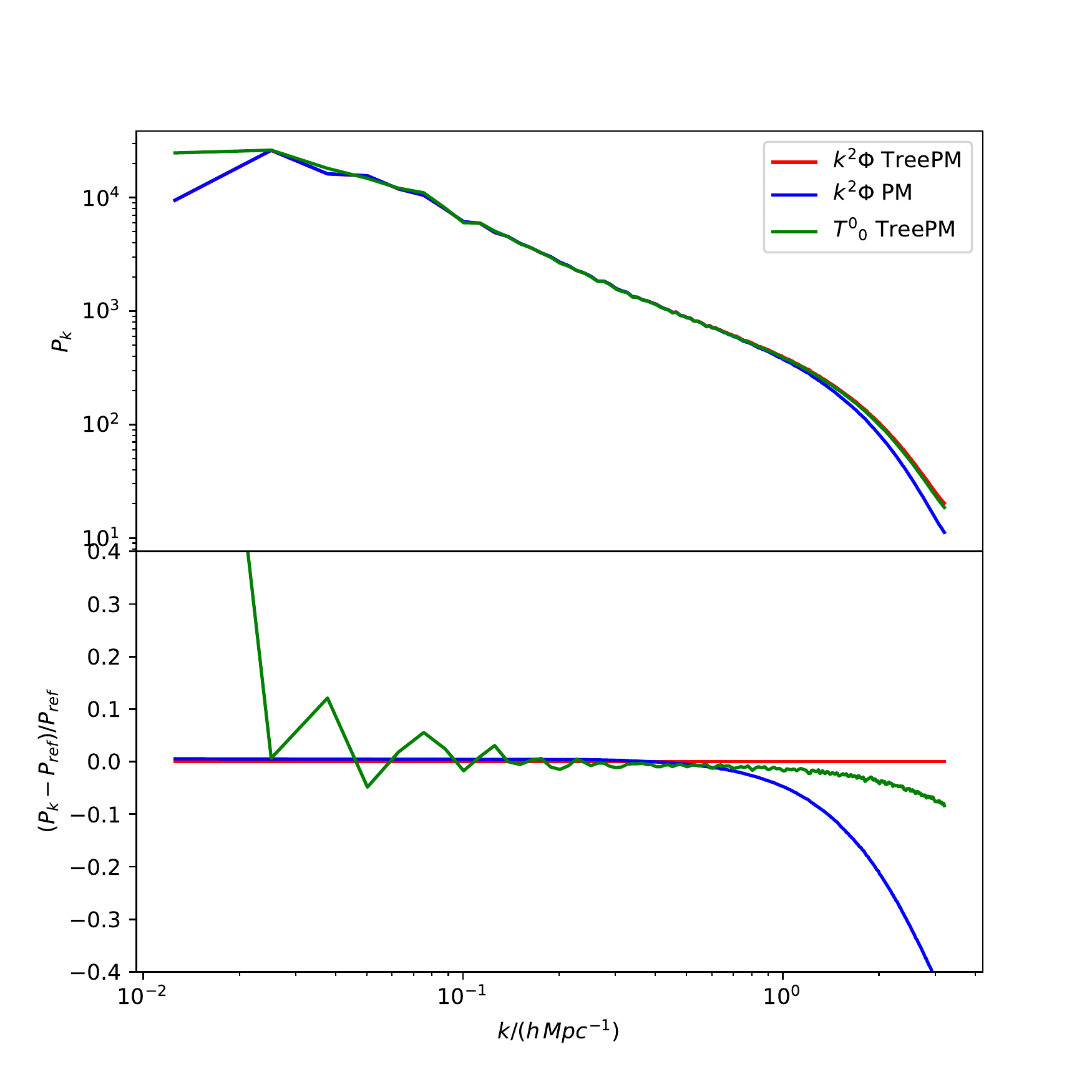}
  \caption{In the left plot: power spectrum of the metric perturbation $\Phi$ 
  in a {\testbig} simulation obtained with
  {\grgadget}.
  In the right plot: power spectrum of $k^2 \Phi$ and $T^0{}_0$.
  For modes well below the Hubble horizon and
  small perturbations it should be verified that $k^2 \tilde \Phi \approx \tilde T^0{}_{0}$.}
  %\pigi{LABELS TROPPO PICCOLE}
  \label{fig:power_phi}
\end{figure*}

\begin{figure*}
    \centering
    \includegraphics[width=.49\textwidth]{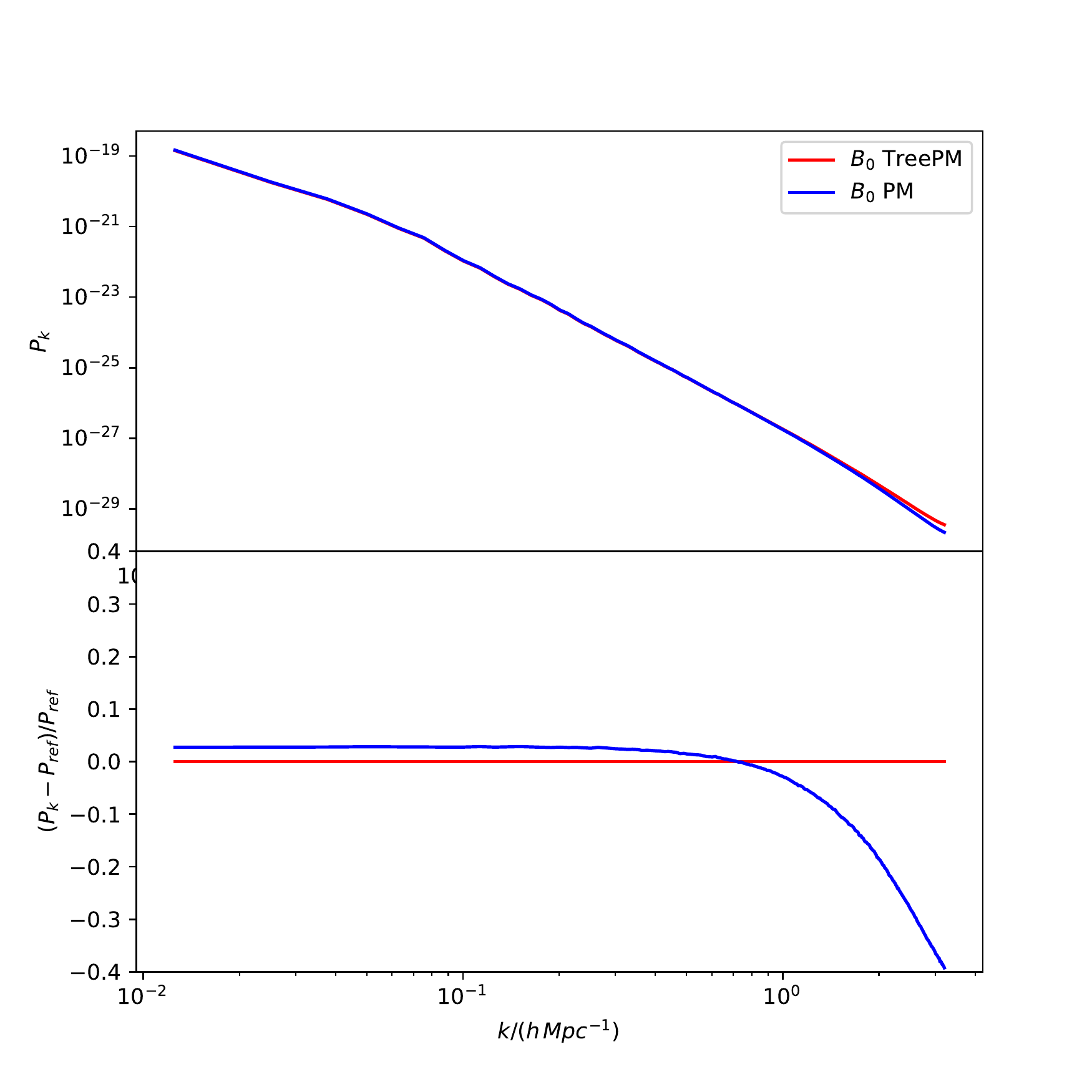}
    \includegraphics[width=.49\textwidth]{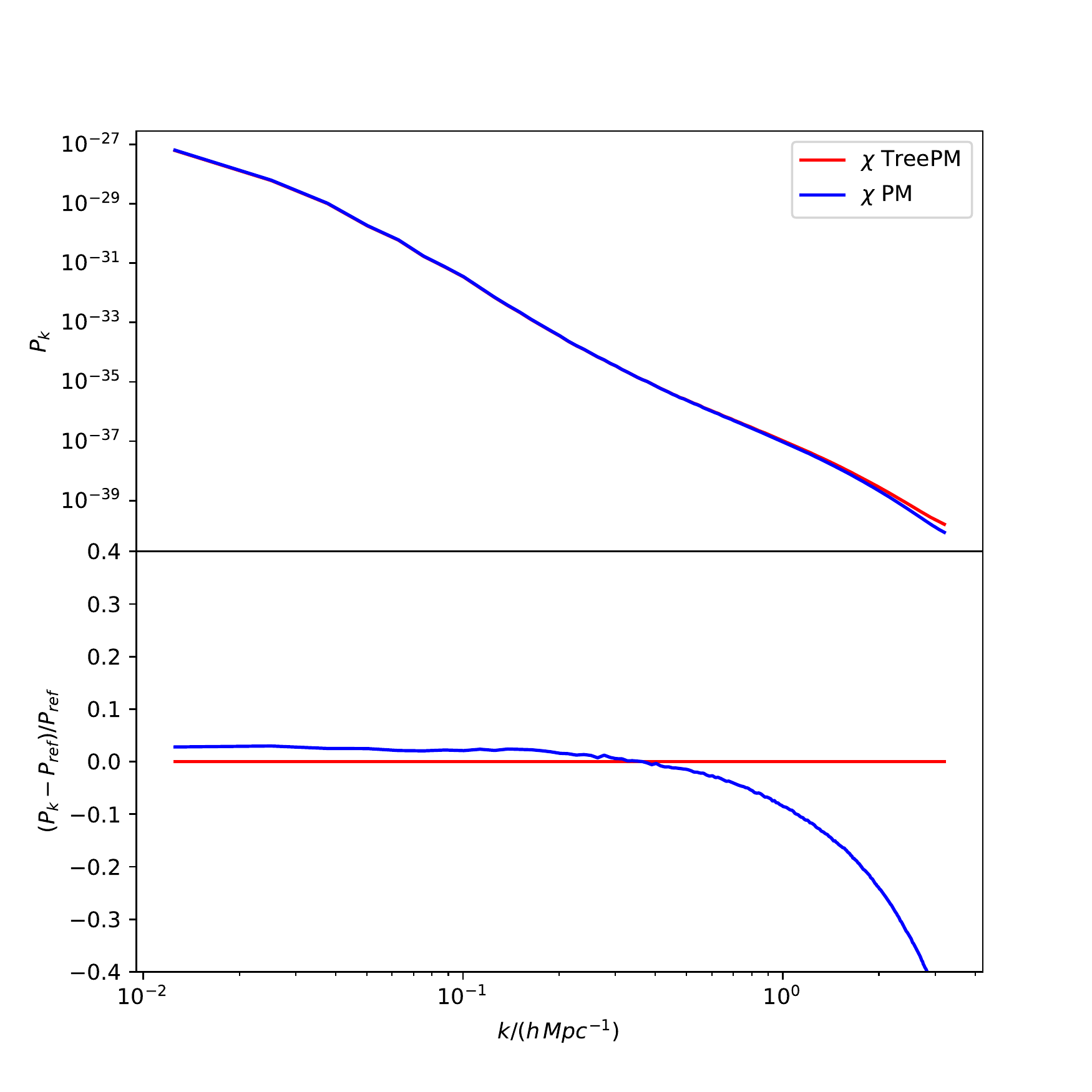}
  \caption{In the left plot: power spectrum 
  of the metric perturbation $B_i$ (the $x$ component) 
  in a {\testbig} simulation obtained with
  {\grgadget}. 
  %\pigi{Media tra le tre componenti?}
  In the right plot: power spectrum of $\chi$.}
  \label{fig:power_Bi}
\end{figure*}

Being designed for the use of Fourier methods from the beginning,
\libgevolution\ offers an interface for the computation of the power spectrum of
the fields defined through the library's interface.
Thus we can also extract and analyse the power spectra of
the individual components of the metric perturbations from the relativistic
simulations.
Figures \ref{fig:power_phi} and \ref{fig:power_Bi}
show the power spectra of the 
relativistic potentials, $\Phi$, $B_i$ and $\chi$, for a high resolution configuration
{\testbig} (see table~\ref{tab:list_configurations}).
These plots show a comparison of {\pmesh} (blue lines) and {\treepm} (red lines) simulations.
The power spectrum of the gravitational potentials converge for both methods
on large scales. However, below $\SI{1}{Mpc}/h$
the {\pmesh}-only simulation loses power with respect to the {\treepm} one;
the differences can reach up to $40\%$ as we approach the Nyquist frequency.
This pattern is equally found for the scalar fields $\Phi$ and $\chi$,
as well as for the individual components of $B_i$.

The right plot in Fig.~\ref{fig:power_phi} helps to understand the reason behind
this result.
Generally speaking, energy density, momentum density and their respective density current (the components of the Energy-Momentum tensor) are sources
of the metric perturbations. Even though those quantities, as fields,
are found at discrete positions of space defined by the mesh,
their values are computed by sampling the energy and momentum carried
by the particle distribution, which contain information on the clustering
due to the short range interactions (through the Tree) that goes well
below the mesh resolution $L/N$. Therefore, {\treepm} simulations, having power on scales well smaller than the PM mesh,
give a better representation of the source of metric perturbation, and thus allow to recover power 
% at the lower frequency modes below Nyquist.
at frequency modes right below Nyquist.
Fig.~\ref{fig:power_phi} highlights 
the particular case of $T^0{}_0$ (the matter density) as a source for $\Phi$;
by comparing $T^0_0$ with $k^2\Phi$, we are verifying the Poisson equation
$k^2 \tilde\Phi \approx\tilde T^0{}_{0}$ that is valid for wavelengths below the Hubble horizon.
%and for small values of $\Phi$  NB: questo non e` vero
This confirms that the presence of small-scale clustering in the particle distribution propagates to the gravitational fields up to the maximum 
resolution that the {\pmesh} allows. The same thing is visible in the vector modes $B_i$ and in $\chi$ (Figure~\ref{fig:power_Bi}), where we also notice a small, few-percent mismatch on large scales. These fields are known to give sub-percent effects on observables, so this difference, that is likely due to some degree of numerical mode coupling, is non considered as a problem.

    % Notice in this case that means that all modes above the number 
    % $ \SI{1}{Mpc}/h \cdot \frac{L}{2\pi} = 80$ approximately are better
    % represented by the \treepm\ simulation
    % and that amounts to $1 - (\frac{80}{256})^3 \approx 0.97$ of all modes.

In figure~\ref{fig:power_gr_cut} we show how the matter power spectrum obtained using {\grgadget} is affected by the choice of the 
gr-smoothing scale parameter $r_b$. We have used an {\testmed} box configuration
to perform this test, and tested values of $r_b = 1.5, 3, 6$ in units of 
$L/N \approx \SI{4}{Mpc}/h$. 
We find that large-scales power is independent of the value of $r_b$ parameter; structures one scales below the {\pmesh} resolution are resolved
by the Tree algorithm, hence for $k > k_{\mathrm{Nyquist}}$ there is a convergence of
all simulations to a common non-linear power spectrum tail.
It is in the medium to small scales $ k_{\mathrm{Nyquist}}> k >\SI{0.2}{Mpc^{-1}}h$ 
that we notice differences in the power spectrum above the $\sim 1\%$ (dashed grey
line).
For small values of $r_b$ ($\sim 1.5\, L/N$),
we obtain discrepancies in the power spectrum at $k\sim \SI{0.5}{Mpc^{-1}} h$
that can be as large as 5 percent and indicate the limitations of our force summation scheme, Eq.~\eqref{eq:force2}. 
A value of $r_b = 3\, L/N$ or possibly higher is needed to obtain a good compatibility 
of {\grgadget} and {\gadget} for all modes greater than $\SI{0.1}{Mpc^{-1}}h$,
where relativistic features in the matter clustering is negligible.

\begin{figure*}
  \centering\includegraphics[width=.9\textwidth]{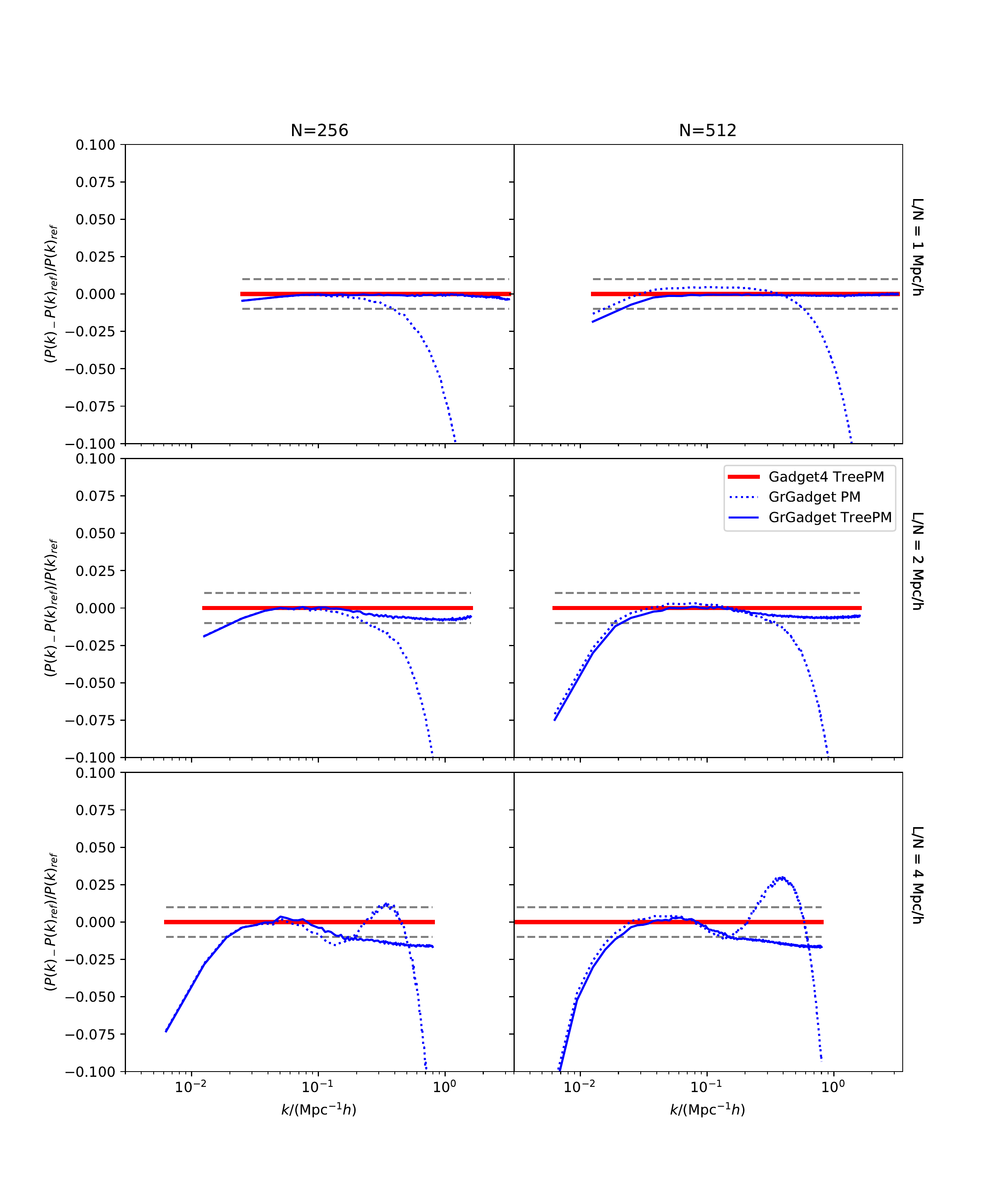}
  \caption{Matter power spectrum from cosmological simulations 
  at $z=0$ using \grgadget\
  (the blue lines)
  and compared to \gadget\ (the red line) at $z=0$. 
  The dotted line is obtained with a simulation in which only the \pmesh\ is
  used to compute forces.
  The plots show the relative difference with respect to the power
  spectrum obtained with \gadget.
  The left column corresponds to simulations
  with $N=256$ grid points per dimension while for the right column $N=512$.
  The boxsize changes along the ranks so that for the top plots the resolution
  is the highest $L/N\approx \SI{1}{Mpc}/h$, in the middle 
  $L/N \approx \SI{2}{Mpc}/h$ and the bottom plots correspond to $L/N\approx
  \SI{4}{Mpc}/h$. In all cases $r_a=1.5\, L/N$ and $r_b = 3\,L/N$.
  The grey dashed line indicate a $1\%$ error.}
  \label{fig:resolution_test}
\end{figure*}

The last test we present here regards the convergence of the numerical results
for increasing resolution.
Figure \ref{fig:resolution_test} shows the matter power spectrum obtained from
running \gadget's \treepm\ (red lines), \grgadget\ with \pmesh-only (blue
dotted lines) and \grgadget\ with \treepm\ (blue continuous line).
These various code configurations were run with different combinations of the
number of grid points per dimension $N=256$, $N=512$ and
box length $L=250$, 500, 1000, $\SI{2000}{Mpc}/h$; the number of particles
was fixed as $N_p = N^3$. In all cases we have set
the {\pmesh} smoothing scale to $r_a=1.5\, L/N$ and 
the gr-smoothing scale to $r_b = 3\,L/N$.
It can be observed with the finest resolution, in the top plots, 
that there is a matching between General Relativity and
Newtonian dynamics in the small scales.
Then as the mesh size becomes coarser, in the middle plots,
some discrepancies in the power spectrum start to appear which 
become more evident for even coarser meshes, in the bottom plots.
This mismatch may be caused by $r_b = 3\, L/N$ 
moving towards larger scales, so that the assumption that PM forces are Newtonian on the small scales breaks. Indeed, while with $L/N=1\ h^{-1}$ Mpc ($r_b=3\ h^{-1}$ Mpc) the scales where relativistic effects become evident in the matter power spectrum and the scales where the pure PM prediction starts to deviate from TreePM are well separated, for larger $L/N$ values the two scales get nearer, indicating that the assumption of pure Newtonian forces on the mesh scale may not be very good. This conclusion is apparently at variance with the discussion of Figure~\ref{fig:power_gr_cut}, where a larger value of $r_b$ was preferred; however, that figure refers to $L/N=1$ and is shown at $z=0.5$, where clustering is a bit weaker.
%forcing the Newtonian limit 
%on the motion equations \eqref{eq:hamilton_eq1} and \eqref{eq:hamilton_eq2}
%at scales where the Hubble horizon cannot be neglected.
%Because we saw already in Fig.~\ref{fig:power_gr_cut} that the contrary 
%effect occurs; i.e. the match between Newtonian simulation and the Relativistic
%one becomes better in the small scale regime when $r_b$ in increased.
%Instead, all combined, these plots put into evidence that the relativistic 
%{\pmesh} becomes numerically unstable and unreliable when it comes to resolving
%forces for modes close the Nyquist frequency for coarse meshes---$10x$ smaller
%in the case of $L/N\approx \SI{4}{Mpc}/h$. Until we investigate further the
s%ource of this behaviour, 
we thus recommend to work with mesh sizes of  $L/N \sim \SI{1}{Mpc}/h$.

% \eduardo{We also notice in the highest resolution run $N=512$, $L=500\SI{}{Mpc}/h$ 
% that the \pmesh-only differs from the rest of the \treepm\ lines at the large
% scales. This is something we need to sort out. I guess this is a problem of
% timestep syncronization with outputs.}

% What did I expect:
% 1. overall matching of GrGadget and Gadget4 at the physical scales 
% for k>0.04 h/Mpc. 
% 2. pushing r_b to close or below r_a will break the TreePM force addition. 

% What we see:
% 1. the matching dependends on the PM resolution.

% The equivalence breaking is due to either:
% 1. just incompatibility of GR and Newtonian gravity in the small scales
% - but when I increase the resolution the two TreePM become compatible at small
% scales even as the k_b (1/r_b) is pushed to higher values. So Newton and GR
% are compatible at small scales.

% 2. consequence of pushing the r_b too far to the large scales (small k) where
% GR should hold (trying to match GR and Newtonian gravity in the large scales)
% - true, as the resolution decreases r_b is pushed to large scales where GR
% becomes more evident. But if we perform a test in which the resolution is
% fixed, we find that the small scale disagreement
% disapears gradually as r_b increases.

% 3. the GR PM has some numerical instability that manifests for low spatial
% resolution.
% - more likely. But why is the numerical stability manifesting with low
% resolution? and what could be the cause of that stability?

{  
  \def\path{images/test-gr-cut}%
  \begin{figure}
    \centering\includegraphics[width=.45\textwidth]{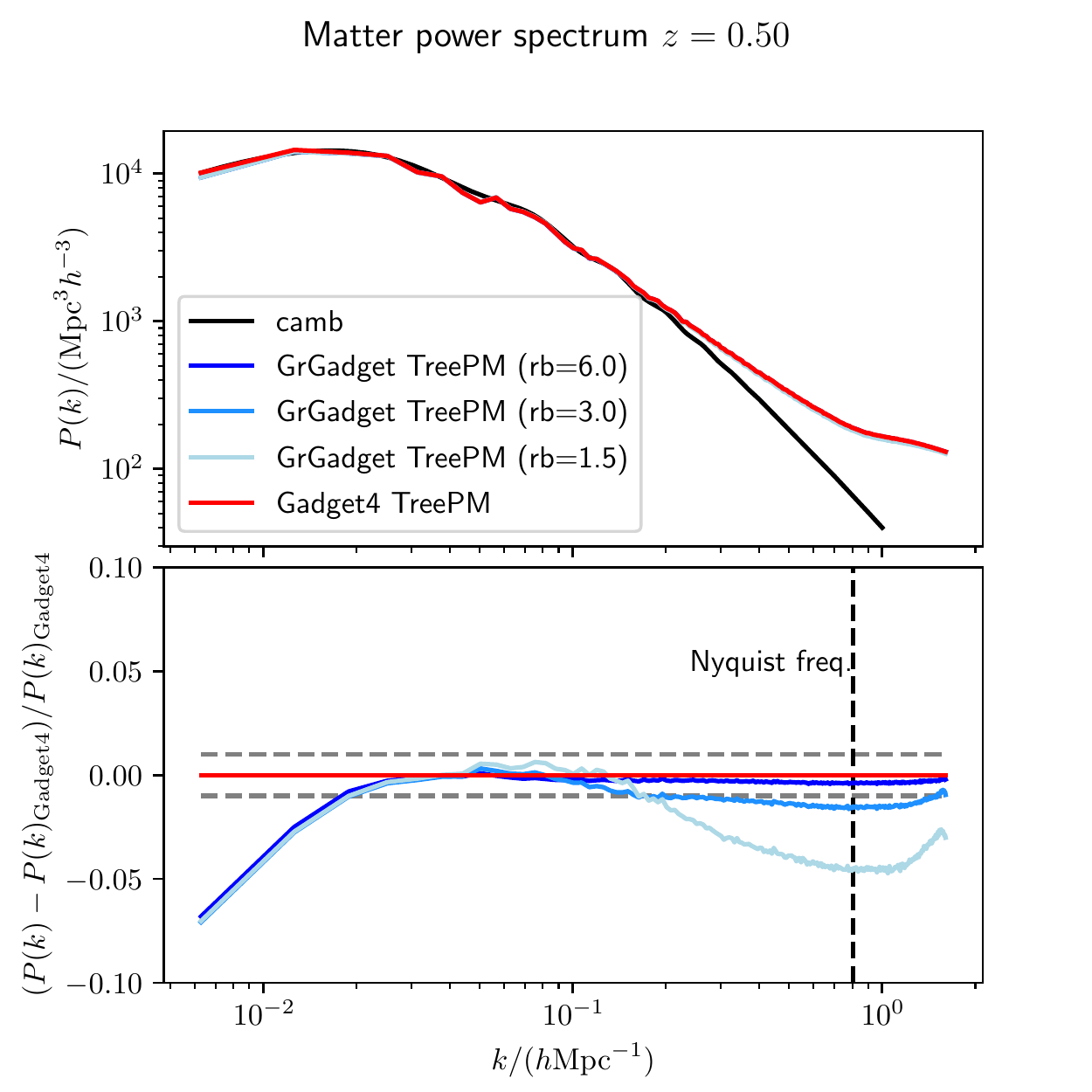}
    \caption{Power spectrum of matter density for
    {\gadget} and {\grgadget}, on a {\testmed} simulation configuration. 
    The upper panel shows the absolute value and the lower panel the relative
    difference with respect to \gadget's \treepm.
    Different shades of blue indicate different values of the gr-smoothing scale
    parameter $r_b=1.5,3,6$ in units of $L/N$.
    The {\pmesh} smoothing scale is $r_a = 1.5\, L/N$.
    The power spectra in this plot are computed beyond the Nyquist frequency
    to show the convergence of the matter distribution correlations for
    distances below the grid resolution, the Tree regime.}
    \label{fig:power_gr_cut}
    % N = 256
    % L = 1000 Mpc/h
  \end{figure}
}

\section{Conclusions}
% \eduardo{Concluding remarks. Is our code working as we expected, yes/no why?
% Does our code brings an improvement to \gadget/\gevolution/to the general
% developement of GR simulations? What can we do with our code to contribute to
% the future survey missions?}

% \eduardo{Our proposal in few words, a TreePM improvement of GR simulations.}
% \eduardo{Advantages of our proposal some possible applications to the context.}

We have constructed a relativistic TreePM code, that we call \grgadget, where the large-scale contribution to the gravitational force is computed using the relativistic C++ \pmesh\ library \libgevolution, based on \gevolution\ code,
while gravity coming from small scales is computed by the Tree code of {\gadget}.
The code works under the assumption that, in the context of cosmological simulations, dark matter can be treated non-relativistically and then the equations of motion of tracer particles tend to the Newtonian limit at scales well below the Hubble horizon.
Following the \gevolution\ approach, we use a weak field approximation of \GR, where the perturbations of the space-time metric with respect to FLRW background are encoded as fields and simulated by the {\pmesh}. 
%The forces acting on the simulation particles are obtained as combination of contributions coming from the relativistic {\pmesh} and the Newtonian short distance gravitational interactions given by {\gadget}'s Tree.
Comparing the matter power spectrum from {\grgadget} simulations with that of original {\gadget} and {\gevolution} codes, we conclude that the code produces consistent results as long as the {\pmesh} cell size $L/N$ is smaller than $\SI{2}{Mpc}/h$
and the gr-smoothing parameter is $r_b \approx 3\, L/N$.

With respect to the pure {\pmesh} implementation of {\gevolution}, the predictive power of {\grgadget} gives an improvement even on the scales sampled by the mesh. This is due to the fact that the energy-momentum tensor, that sources the equations of the fields that represent the perturbations of the metric, is computed from a fully non-linear distribution of particles, with gravity being resolved down to a much smaller softening length and not down to the mesh size. This may be very useful, e.g., when assessing the possibility of detecting the frame-dragging effect of a rotating dark-matter halo, if not of a spiral galaxy \citep{Bruni2014}.
%By representing power on scales much smaller than a fixed mesh, {\grgadget} brings an improvement to {\gevolution} in that the source terms for the fields that describe the perturbation of the metric correctly take into account the highly non-linear nature of the matter field below the mesh scale. 
%It brings an improvement to \gadget, by giving a relativistic treatment in the limit (scales not much smaller than the horizon, relativistic particles having little power on small scales due to free streaming) that is mostly relevant to cosmological analyses.
%While its {\treepm} method is able to resolve a high dynamic range in
%spatial resolution so that the formation of highly non-linear structures in the
%matter distribution can be obtained even in large cosmological volumes.
Furthermore, this code is a development of the widely used {\gadget} code, and because the PM sector of the code is called only by the computation of the gravity force, our code can be easily extended to simulations of galaxies or galaxy clusters by switching on the hydrodynamics, star formation and feedback sectors. All the physics described by these sectors can safely be treated in the Newtownian limit; one should in principle add thermal energy of gas particles to the energy-momentum tensor, but while this extension is straightforward, it is likely to provide a negligible contribution.

This is, for our group, a further step in the construction of an ecosystem of simulation codes and post-processing tools for modeling the evolution of structure in the Universe, with the aim of making predictions for precision cosmology. Sub-percent accuracy in cosmological predictions, that matches the smallness of the statistical error that will be obtained with forthcoming galaxy surveys mentioned in the Introduction, can only be obtained taking into account relativistic effect \citep[e.g.][]{Lepori2020}, and we can foresee that a self-consistent treatment of these effects (to within the required accuracy) will soon become the standard in cosmological simulations. These effects can also be added by post-processing Newtonian simulations, but a validation of these procedures requires validation against a more self-consistent approach. Conversely, a large community is developing {\gevolution} in the direction of adding modifications of gravity, whose formulation is typically worked out in a general relativistic context. This line of development, coupled with a Newtonian treatment of modified gravity in the Tree code, would be precious in the formulation of tests of gravity, because relativistic effects may hide smoking-gun features of specific classes of modified gravity theories.

\appendix
\section{Code scaling}
\label{sec:code_scaling}

The code we presented in this work is the merging of two codes whose behaviour in terms of run-time scaling is well-known and characterized; since we did not modify the underlying algorithms, it is expected that the run-time scaling of our code follows that of the parent codes.

However, the {\libgevolution}'s PM is obviously different from {\gadget}'s, and we added the translation of particles data from the host code to the target relativistic PM. Both this facts require that we establish the overall scaling of {\grgadget} in its fully-relativistic configuration and the overhead associated to both the relativistic PM and the interface between the two codes.

\begin{figure}
    \centering
    \includegraphics[width=.45\textwidth]{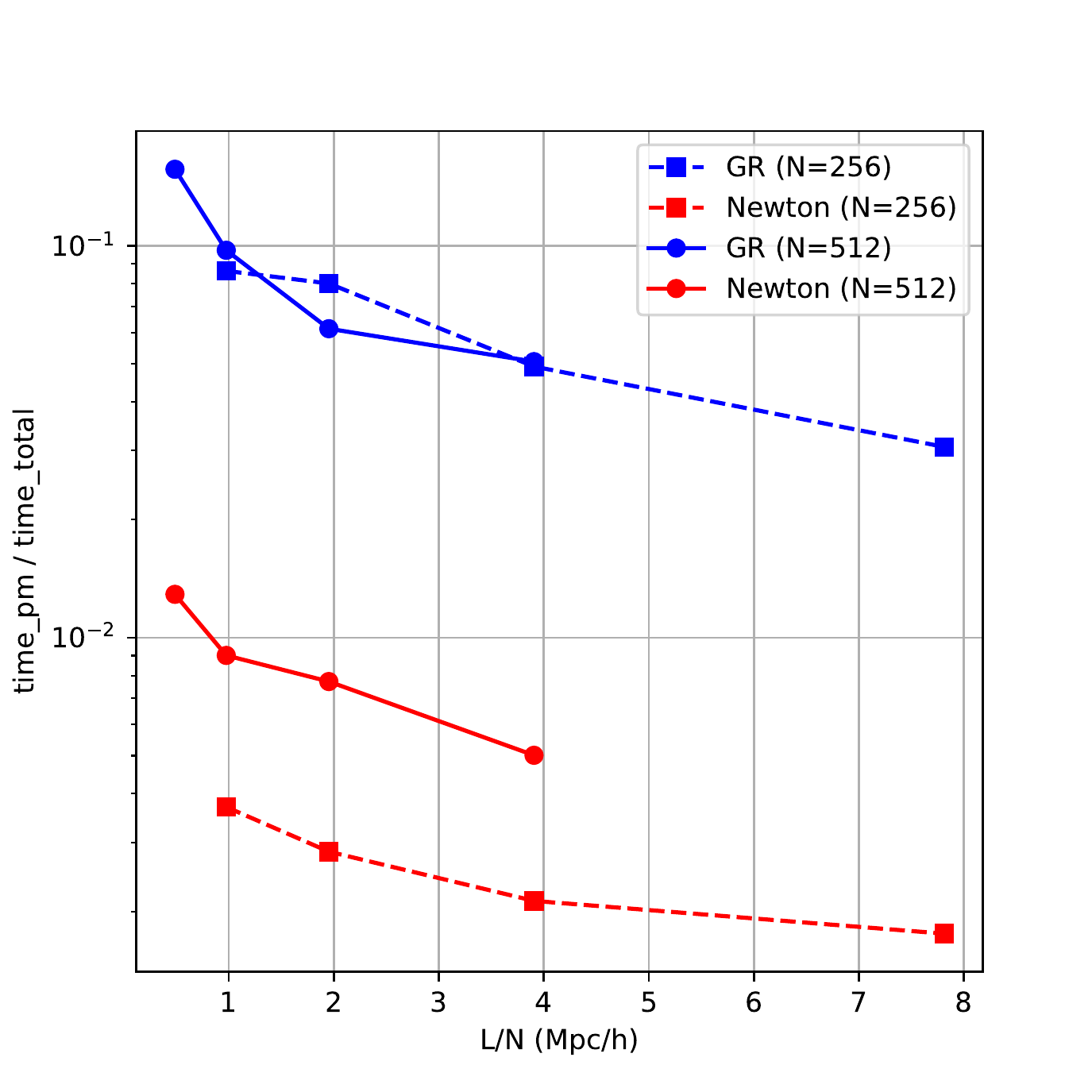}
    \caption{The fraction of PM time to the total running time. Relativistic runs are shown in blue while Newtonian runs are shown in red, whereas symbols distinguish the value of grid points per dimension $N$ (squares and circles for $N=256$ and $512$ respectively). We plot the time fraction on the $y$--axis (log scale) vs the mesh resolution $L/N$ on the $x$--axis.}
    \label{fig:pm_overhead}
\end{figure}

In figure \ref{fig:pm_overhead} we show the fraction of time spent in the PM in both the original and relativistic configurations as a function of the grid cell size (see the caption for details). The relativistic PM is an order of magnitude more expensive than the original $\gadget$'s Newtonian PM, although in absolute sense it is still either negligible or secondary in the simulation sets that have been tested (it reaches a maximum value of $16\%$ at highest resolution, i.e. in the $N=512$,$L=\SI{250}{Mpc}/h$). However, it scales with both the resolution and the grid number as the original Newtonian PM does.

% Eduardo: We don't know this.
%The time spent in the particles data translation between the two codes results to be %completely negligible.

\begin{figure*}
    \centering
    \includegraphics[width=.33\textwidth]{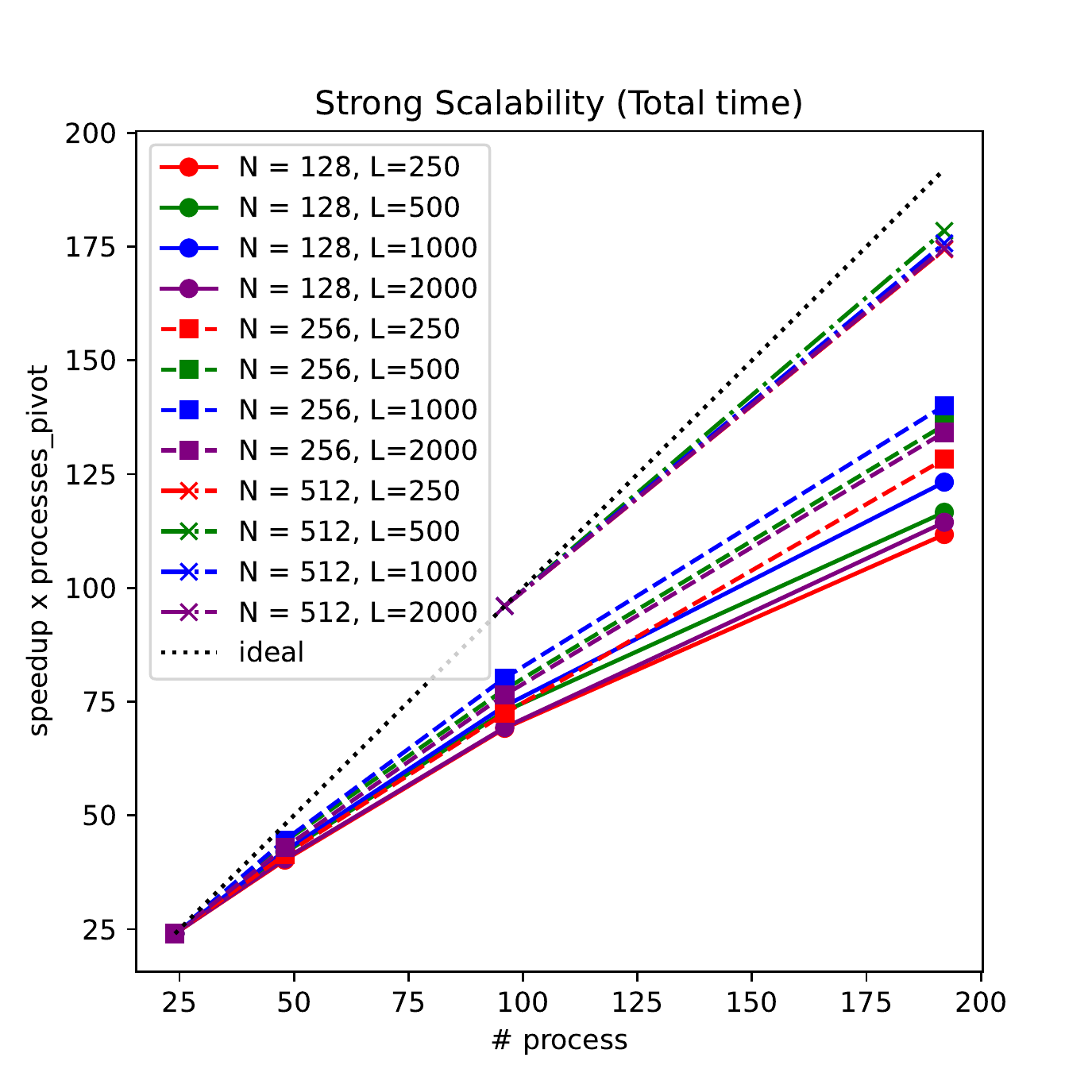}
    \includegraphics[width=.33\textwidth]{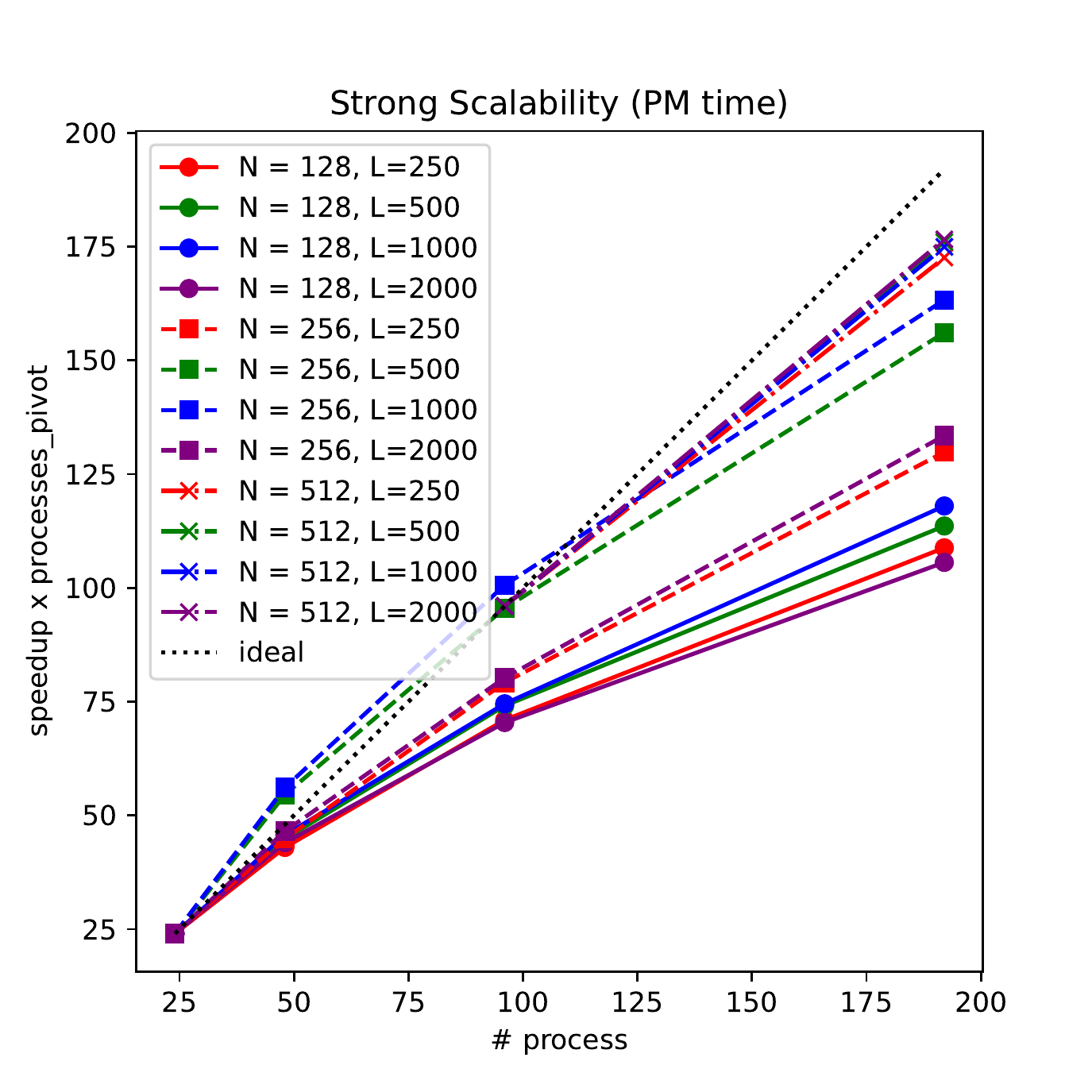}
    \includegraphics[width=.33\textwidth]{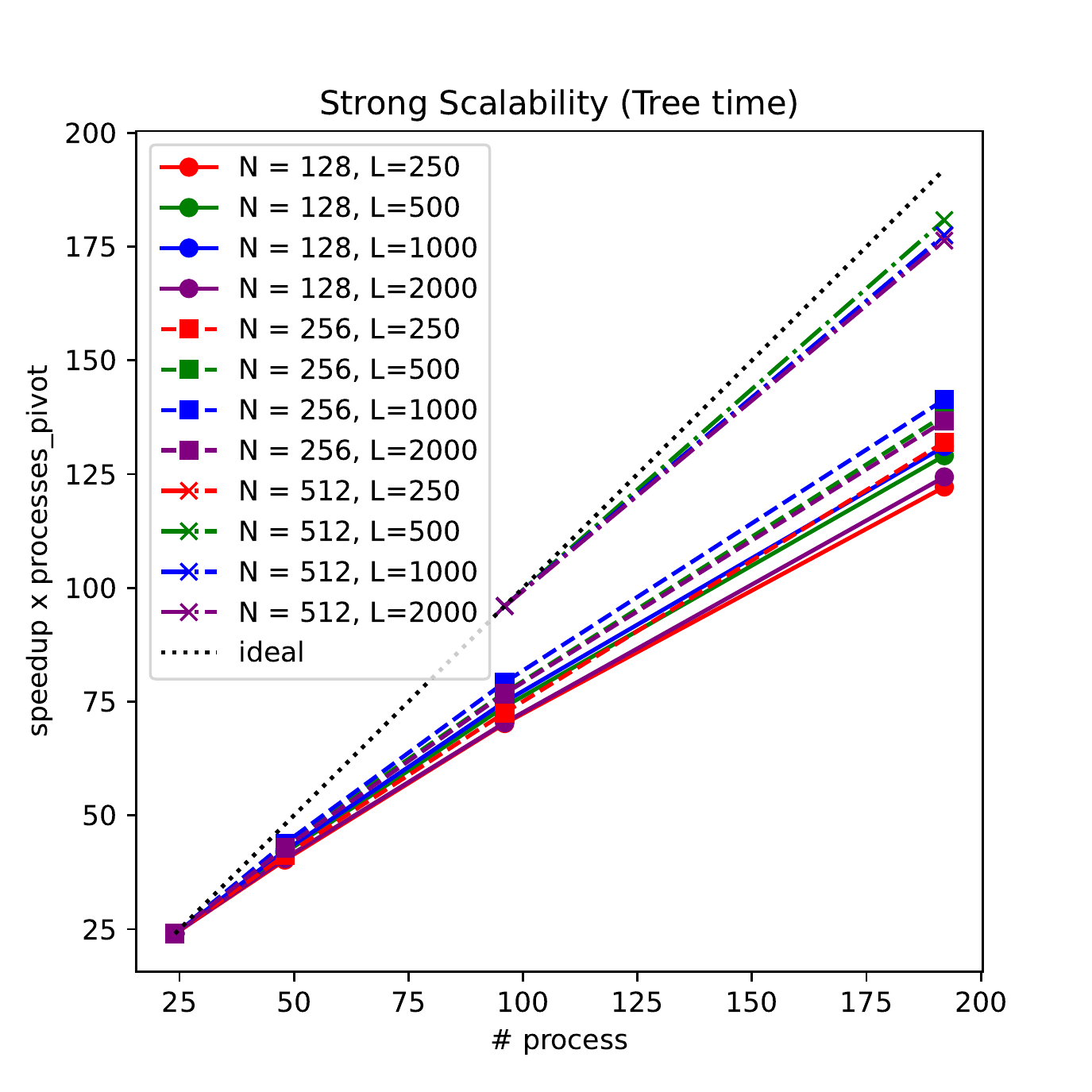}
    \caption{Strong-scaling test. We present the code scaling as the number $P$ of MPI tasks is increased while running the same simulation set-up. All the results refer to {\grgadget}, i.e. to the configuration with fully-relativistic PM. On the $x$--axis $P$ increases from 24 to 192, by $\times 2$ steps. On the $y$--axis we report the speed-up (normalized so that the ideal speed-up for $P=1$ is 1)
    %@Eduardo: why is the timing at #process = 24 also equal to 24? shouldn't it be 1 ?
     for the total running time, the time spent in the PM and the time spent in the Tree on the Left, Middle and Right panels respectively. Note that the ideal behaviour (black dotted line) would result in a linear speed-up. The PM data includes the translation of particles data from {\gadget} to {\libgevolution}.
     We show the results for $N=128$, $256$ and $512$ (solid, dashed and dot--dashed lines respectively) for 4 different box sizes (i.e. mass resolutions), $L=250$, $500$, $1000$ and $\SI{2000}{Mpc}/h$ (circles, squares and stars respectively).
     See the discussion in Appendix \ref{sec:code_scaling} for details.
    }
    \label{fig:strong_scaling}
\end{figure*}

\begin{figure*}
    \centering
    \includegraphics[width=.33\textwidth]{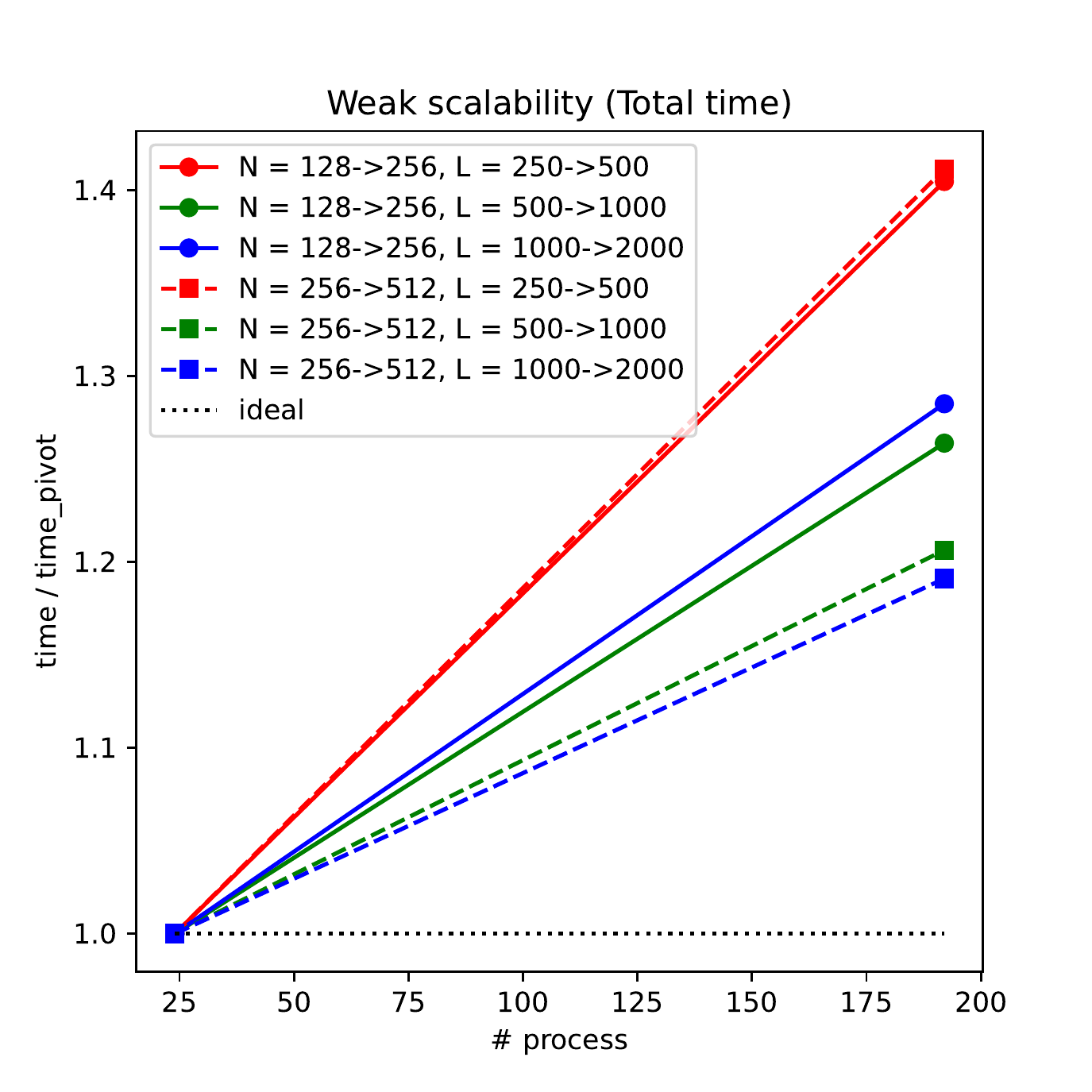}
    \includegraphics[width=.33\textwidth]{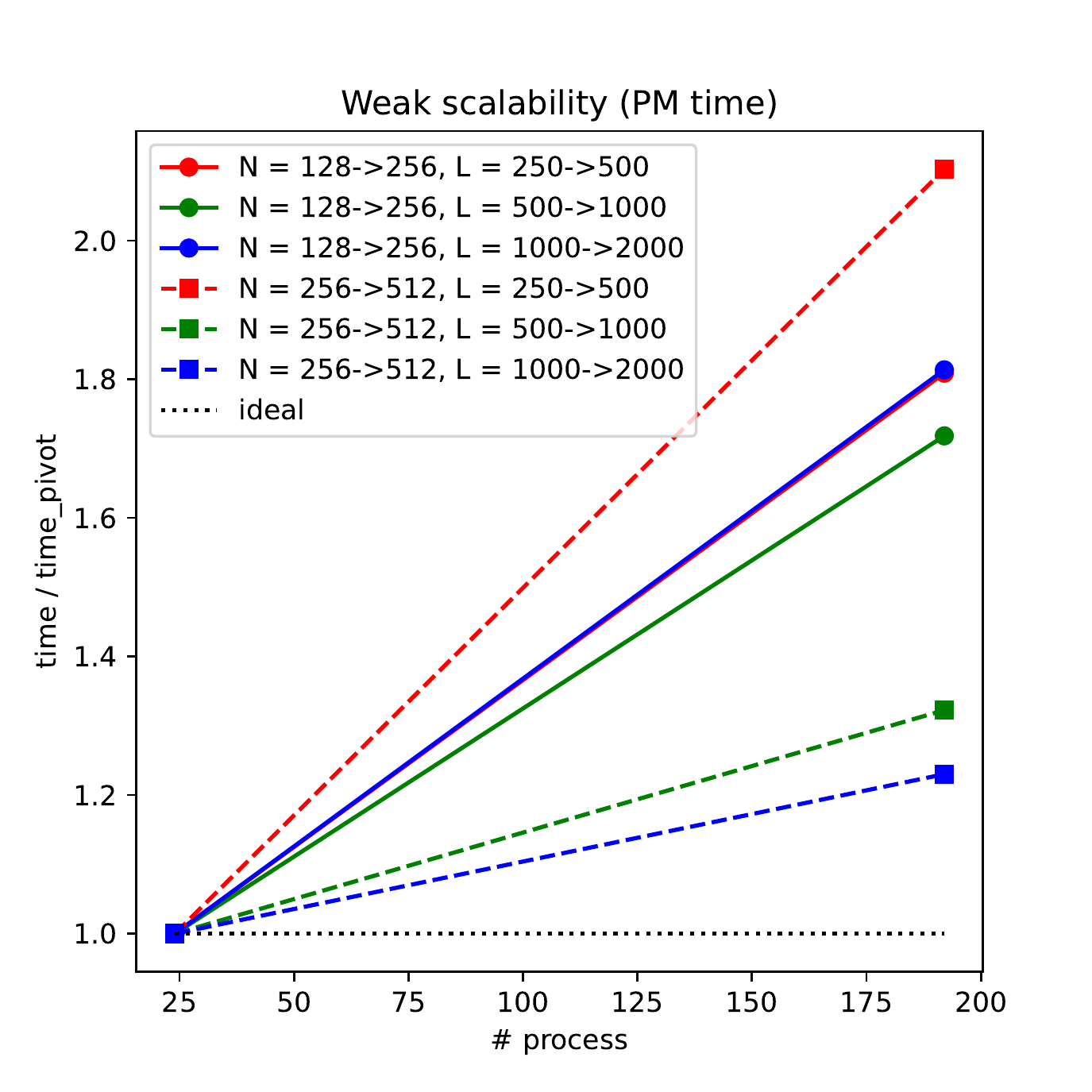}
    \includegraphics[width=.33\textwidth]{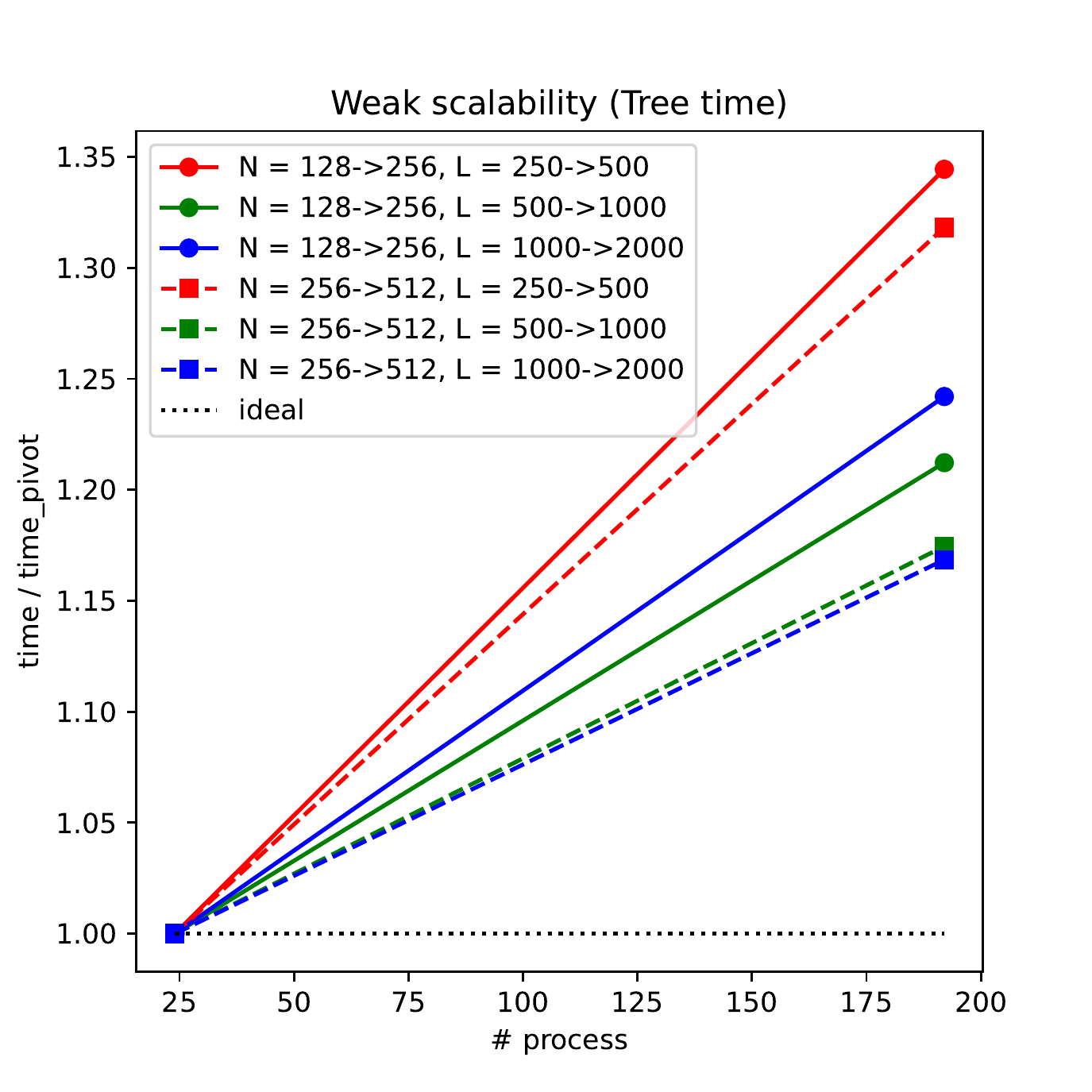}
    \caption{Weak-scaling test. We present the code scaling as the number $P$ of MPI tasks is increased for a proportionally increasing problem, then keeping constant the particles--per--task occupancy. All the results refer to {\grgadget}, i.e. to the configuration with fully-relativistic PM. On the $x$--axis $P$ increases from 24 to 192 with only 2 test cases. On the $y$--axis we report the speed-up
     for the total running time, the time spent in the PM and the time spent in the Tree on the Left, Middle and Right panels respectively. Note that the ideal behaviour would result in a constant running time (horizontal dotted black line). The PM data includes the translation of particles data from {\gadget} to {\libgevolution}.
     We show the results for two cases: from $N=128$, to $N=256$ (solid lines with circles), and from $N=256$, to $N=512$ (dashed lines with squares). Each of the two cases has been run for three different box sizes (i.e. mass resolutions): $L=250\rightarrow L=\SI{500}{Mpc}/h$, $L=500\rightarrow L=\SI{1000}{Mpc}/h$ and $L=1000\rightarrow L=\SI{2000}{Mpc}/h$ (red, green and blue colors respectively).
     See the discussion in Appendix \ref{sec:code_scaling} for details.
      }
    \label{fig:weak_scaling}
\end{figure*}

Figures \ref{fig:strong_scaling} and \ref{fig:weak_scaling} report the scaling of run time in strong and weak scaling tests respectively for the total run time, the tree time and the PM time (left. middle and right panels in both figures; see the captions for details).
As inferred from \ref{fig:pm_overhead}, the run-time and hence its scaling, are dominated by the {\gadget}'s Tree section. 
%{\bf TO BE CONCLUDED.. just few more lines}


\section*{Acknowledgements}

We thank Julian Adamek for many fruitful discussions on {\sc gevolution}, Volker Springel for his comments on an early draft, Francesca Lepori, Marco Bruni, Marco Baldi and Emilio Bellini for discussions.
Simulations were performed with the HOTCAT system of INAF \citep{Taffoni2020,Bertocco2020}. PM acknowledges partial support by a {\em Fondo di Ricerca di Ateneo} grant of University of Trieste.

%The Acknowledgements section is not numbered. Here you can thank helpful
%colleagues, acknowledge funding agencies, telescopes and facilities used etc.

%%%%%%%%%%%%%%%%%%%%%%%%%%%%%%%%%%%%%%%%%%%%%%%%%%
\section*{Data Availability}

The simulation codes presented in this paper are publicly available on github
in the following path:
\url{https://github.com/GrGadget}.

\bibliographystyle{mnras}
\bibliography{mybibtex}

\appendix

\bsp % typesetting comment
\label{lastpage}
\end{document}